\documentclass[11pt, letterpaper]{article}
\usepackage[margin=.9in]{geometry}
\usepackage[utf8]{inputenc}
\usepackage{graphicx}
\usepackage{float}
\usepackage{color}
\usepackage{amsmath, amsthm, amssymb, amsfonts, mathtools}
\usepackage[round]{natbib}
\usepackage{hyperref}
\usepackage{enumitem}
\usepackage[bottom, ragged]{footmisc}
\usepackage{caption}
\usepackage{subcaption}

\numberwithin{equation}{section}

\definecolor{Blue}{RGB}{26, 110, 178}
\definecolor{Orange}{RGB}{250, 120, 10}

\usepackage{hyperref}
\hypersetup{
        colorlinks=true, 
        linkcolor=Orange,
        citecolor=Blue,
        urlcolor=Blue,
        pdfborder={0 0 0},
}

\newtheorem{definition}{Definition}
\newtheorem{proposition}{Proposition}
\newtheorem{lemma}{Lemma}

\usepackage[linesnumbered, ruled, vlined]{algorithm2e}

\vspace{12pt}
\title{\textsc{
Attention Elasticities\\
and Invariant Information Costs\thanks{I thank Andrew Caplin for helpful comments and discussions.}
}}
\author{D\'{a}niel Csaba\thanks{QuantCo. Email: \href{mailto:csaba.daniel@gmail.com}{csaba.daniel@gmail.com}}}
\date{April, 2021}

\begin{document}

\maketitle

\vspace{74pt}
\begin{abstract}
We consider a generalization of rational inattention problems by measuring costs of information through the information radius \citep{sibson1969information, verdu2015alpha} of statistical experiments. We introduce a notion of attention elasticity measuring the sensitivity of attention strategies with respect to changes in incentives. We show how the introduced class of cost functions controls attention elasticities while the Shannon model restricts attention elasticity to be unity. We explore further differences and similarities relative to the Shannon model in relation to invariance, posterior separability, consideration sets, and the ability to learn events with certainty. Lastly, we provide an efficient alternating minimization method---analogous to the Blahut-Arimoto algorithm---to obtain optimal attention strategies.
\vspace{3pt}

\end{abstract}

\setlength{\parindent}{0pt}
\setlength\parskip{9pt}

\newpage
\section{Introduction}

The pivotal role of information has long been recognized in economic decisions. Understanding the constraints that shape the beliefs decision makers hold helps us understand and predict their behavior in different environments. The model of rational inattention \citep{sims1998,sims2003} has become a widely adopted framework for the analysis of beliefs and behavior when constraints on attention and information acquisition are of primary importance.\footnote{For an extensive overview of the rational inattention literature see \cite*{mackowiak2020rational}.}

A central question of the rational inattention (RI) literature has been identifying and understanding the behavioral implications of different classes of information cost functions. Our assumptions on the shape of information costs fully determine patterns in behavior under the assumption of Bayesian rationality and allowing unrestricted preferences.

In the present paper we introduce a notion of \textit{attention elasticity}, a simple behavioral implication of information costs. Attention elasticity captures how sensitively attention strategies change with changes in incentives. One expects that increasing the payoff-difference between two actions in a given state increases the conditional odds of the better action being chosen as the decision maker pays more attention. Similarly, we expect that such a local change in payoff-differences is also reflected on the unconditional odds. The posterior odds links these quantities and is equal to the ratio of the conditional odds over the unconditional odds. Our notion of attention elasticity captures the relative rate of change in posterior odds as we change incentives.

We show that mutual information costs imply an attention elasticity of unity irrespective of the payoff structure and prior beliefs. This puts significant restrictions on the types of beliefs and choice probabilities that one can attain within the Shannon model as introduced in the seminal works of \cite{sims1998,sims2003} and widely used in applications. We show that such patterns in attention elasticity can be understood through a perspective of the mutual information as the average divergence of signal probabilities relative to the barycenter of the Blackwell experiment. This motivates our generalization of the Shannon model.

In particular, we analyze a novel class of information cost functions that relaxes the restrictive elasticity patterns of the Shannon model while maintaining many of its desirable features. The introduced class generalizes mutual information by characterizing costs of statistical experiments based on the average R\'enyi divergence of conditional signal probabilities relative to the barycenter of the experiment \citep{sibson1969information,verdu2015alpha}, a class named $\alpha$-mutual information.\footnote{The recent work of \cite*{hoiles2020rationally} in the computer science literature has used the same class of information costs in an inverse reinforcement learning task without exploring behavioral implications.} We show that such a view gives rise to information costs that are analogous to constant elasticity of substitution production functions widely used in other domains of economics. This gives clear behavioral meaning to information cost parameters affecting the changes in beliefs and behavior due to changes in incentives. We refer to the corresponding class as $\alpha$-rational inattention ($\alpha$-RI). In general symmetric decision problems we provide conditions under which attention strategies exhibit elastic or inelastic patterns in $\alpha$-RI problems.

We show that $\alpha$-mutual information satisfies invariance but is not posterior separable providing a novel direction for generalizing the Shannon model. Invariance can be a desirable property in several economic contexts. For instance, in large games with incomplete information invariance of information costs is a prerequisite for the efficiency of equilibria and it prevents non-fundamental volatility as shown by the recent work of \cite{hebert2020information} and \cite{angeletos2019inattentive}.

Our approach thus complements several recent papers which have provided generalizations of the Shannon model that result in various posterior separable classes of information cost functions. \cite*{caplin2019rationally} show that mutual information is uniquely characterized within the class of uniformly posterior separable cost functions by the invariance under compression property, which is a general implication of invariant cost functions. On the other hand, invariance implies a strong form of symmetry across states and its implications can be restrictive in other contexts. Neighborhood-based cost functions, such as the ones proposed by \cite{hebert2020neighborhood}, and \cite*{pomatto2019cost}, aim to accommodate scenarios in which the costs of learning are asymmetric across states. Posterior separable cost functions are natural in sequential learning problems as shown by \cite{morris2017wald}, \cite{hebert2019rational}, and \cite{bloedel2020cost} for instance.

We explore further behavioral implications. We show that under $\alpha$-mutual information costs the rationally inattentive decision maker might learn payoff-relevant events with certainty, a behavior that is impossible in the Shannon model. We also characterize consideration sets and provide a generalization of the conditions corresponding to the Shannon case \citep*{caplin2019rational}.

Lastly, the introduced class of rational inattention problems are tractable and can be efficiently solved through a generalized Blahut-Arimoto-type algorithm, hence have great appeal for empirical applications. We provide guarantees for convergence and discuss details of implementation.

The rest of the paper is organized as follows. Section \ref{sec: att_elast} introduces the notion of attention elasticity and discusses elasticity patterns under the Shannon model. Section \ref{sec: info_rad} introduces $\alpha$-mutual information costs and presents its main properties. Section \ref{sec: alpha_RI} derives elasticity patterns in $\alpha$-RI problems. Section \ref{sec: invariance} discusses the invariance properties of $\alpha$-mutual information and its corresponding behavioral implications. Section \ref{sec: ucc} shows that under the introduced class, learning payoff-relevant events with certainty is possible. Section \ref{sec: consideration} characterizes consideration sets. Section \ref{sec: mod_BA} presents a modification of the Blahut-Arimoto algorithm providing efficient solutions for optimal attention strategies in $\alpha$-RI problems. Section \ref{sec: conclusion} concludes.

\section{Attention Elasticities and the Shannon Model}
\label{sec: att_elast}


We consider general rational inattention problems with finite domains. Specifically, we consider decision problems described by a finite set of actions, $a\in\mathcal{A}$, and a finite set of states, $\omega\in\Omega$. The prior belief over states is denoted $\mu\in\Delta(\Omega)$, and the utility function is denoted by $u\colon \mathcal{A}\times\Omega \to \mathbb{R}$. The decision maker (DM) can acquire information and base her action on the realization of a signal from a statistical experiment. A statistical experiment is a Markov kernel, $P\colon \Omega\to \Delta(\mathcal{S})$, defining a collection of probability distributions over signals $s\in\mathcal{S}$ for each realization of the state. The conditional signal distributions are denoted as $P(\cdot\mid\omega)$ or as $P_\omega$. A simple experiment has a finite signal support. Keeping the state space fixed, we denote the set of simple experiments as $\mathcal{P}$. For a given experiment, $P$, and prior distribution over states, $\mu$, denote the joint distribution as $\mu  P\in\Delta\left(\Omega\times\mathcal{S}\right)$; the marginal distribution over signals---derived from the joint distribution $\mu P$---as $\mu P_{\mathcal{S}}$; and the joint distribution obtained from the product of two marginals, $\mu\in\Delta(\Omega)$ and $q\in\Delta(\mathcal{S})$, as $\mu \otimes q$.

We allow information costs to depend on the statistical experiment and the prior, and denote $K \colon \mathcal{P} \times \Delta(\Omega) \to \overline{\mathbb{R}}$. We only consider information cost functions that satisfy the convexity property such that, without loss of generality, we can identify the set of signals with the set of actions, $\mathcal{S}\equiv\mathcal{A}$, and use the two sets interchangeably.

The problem of the rationally inattentive DM is then to choose a statistical experiment that maximizes expected utility net of information costs,\footnote{We assume $K$ is proper and lower-semicontinuous so the optimum is attained.}
\begin{equation}
\max_{P\in\mathcal{P}}\ \sum_{\omega, s}\mu(\omega)P(s\mid\omega)u(s,\omega) - K(P, \mu).\
\label{eq: RI_problem}
\end{equation}

Following the seminal works of \cite{sims1998,sims2003} mutual information has been the most widely used information cost in the RI literature---for a large part due to its tractability. The mutual information is defined as the Kullback-Leibler (KL) divergence between the joint distribution and the product distribution derived from the marginals,
\begin{equation}
I(P, \mu) := D\left(\mu P \Vert \mu \otimes \mu P_\mathcal{S}\right) := \sum_{\omega, s} P(s\mid \omega)\mu(\omega) \log \frac{P(s\mid \omega)\mu(\omega)}{\mu P_\mathcal{S}(s)\mu(\omega)}. \label{eq: mi}
\end{equation}
We convert units of information costs (nats) to units of expected utility through a generic scalar, $\kappa$, to obtain
\begin{equation}
    K(P, \mu) = \kappa I(P, \mu).
\end{equation}
With these preliminaries in hand we introduce our notion of attention elasticity.

\subsection{Attention Elasticity}

The attention strategy of the DM is described by the statistical experiment, $P^*$, the solution to the RI problem \eqref{eq: RI_problem}. This specifies the conditional probability of choosing a certain action for each possible realization of the state. Loosely speaking, we expect that if stakes are higher the inattentive DM pays more attention and chooses the action that maximizes her utility with higher probability for each realization of the state.

Following this reasoning, a fundamental characteristic of attention strategies is how the conditional odds of choosing action $a$ over $b$, $P^*(a\mid\omega)/P^*(b\mid\omega)$, change as we change corresponding payoff differences, $u(a, \omega) - u(b, \omega)$. Such local changes in payoff differences are also reflected in changes on the unconditional odds, $\mu P_{\mathcal{S}}^*(a)/\mu P_{\mathcal{S}}^*(b)$. The posterior odds induced by the optimal experiment links these quantities and is given by the ratio of conditional and unconditional odds,
\begin{equation}
    \frac{\gamma^a(\omega)}{\gamma^b(\omega)} = \frac{P^*(a\mid\omega)}{P^*(b\mid\omega)}\frac{\mu P_{\mathcal{S}}^*(b)}{\mu P_{\mathcal{S}}^*(a)}.
\end{equation}

This connects to an equivalent way of modeling attention strategies through distributions over posteriors such that the posteriors satisfy the martingale property of averaging back to the prior. We capture the local sensitivity of attention strategies by analyzing the sensitivity of posterior odds with changes in incentives. We introduce our notion of attention elasticity under arbitrary changes to the payoff structure.

\begin{definition}[Attention elasticity]
    The attention elasticity between actions $a$ and $b$ in state $\omega$ is defined as the relative rate of change in posterior odds as we change incentives. Letting the posterior odds be denoted by
    \begin{equation}
        \Gamma^{a, b}_\omega := \frac{\gamma^a(\omega)}{\gamma^b(\omega)} = \frac{P^*(a\mid\omega)}{ \mu P^*_{\mathcal{S}}(a)}\frac{\mu P^*_{\mathcal{S}}(b)}{P^*(b\mid\omega)},
    \end{equation}
    and letting $\theta$ be a parameter of the incentive structure the attention elasticity is given by,
    \begin{equation}
        \epsilon^{a, b}_{\omega, \theta} := \frac{\partial \Gamma^{a, b}_\omega}{\partial \theta} \frac{1}{\Gamma^{a, b}_\omega}.
    \end{equation}
    We consider only $a, b$ and $\omega$ such that conditional choice probabilities are interior, $P^*\left(a\mid\omega\right), \linebreak P^*\left(b\mid\omega\right)>0$.
\end{definition}

The relative rate of change in the posterior odds is equal to the relative rate of change in conditional odds minus that in unconditional odds. Under certain parametrization of the incentive structure, changes in the incentives affecting all actions leave unconditional choice probabilities intact, only affecting conditional choice probabilities. In such cases our notion of attention elasticity is equivalent to the relative rate of change in conditional choice probability odds.

Given a fixed information cost function and a prior the attention strategy is dependent only on the differences in payoffs per units of information costs of the form, $(u(a, \omega) - u(b, \omega))/\kappa$. Any additive constant applied uniformly for each state-dependent payoff has no impact on the attention strategy. Correspondingly, we will consider changes in the incentive structure by changing  absolute payoff differences per units of information costs across some actions. A scenario of key interest is in which we change the payoff per unit of information cost of a single action, $\theta = u(a, \omega)/\kappa$, and analyze the changes in the corresponding posterior ratio, $\Gamma^{a, b}_\omega$.

It is worth noting the similarities in our definition of attention elasticity and that of elasticity of substitution applied in the context of a profit-maximizing firm. The optimal production plan of the profit-maximizing firm is only dependent on relative price differences across available inputs and the price of the output. We correspondingly analyze the relative rate of change in input ratios with relative changes in price ratios. On the other hand, the attention strategy of a rationally inattentive DM is only dependent on absolute payoff differences per units of information costs. Correspondingly we analyze the relative rate of change in posterior odds with changes in absolute payoff differences.


The Shannon model puts stringent behavioral restrictions on attention elasticities. While immediate we state it as a proposition to emphasize its importance.

\begin{proposition}
    \label{prop: elast_shannon}
     Under Shannon mutual information costs, for any two actions, $a, b$, that are chosen with strictly positive probability under the optimal attention strategy of the rationally inattentive DM, the attention elasticity in state $\omega$ with $\theta = u(a, \omega)/\kappa$ is identically equal to one.
    \begin{equation}
        \epsilon^{a, b}_{\omega, \theta} = 1 \qquad \forall\omega\in\Omega.
    \end{equation}
    For any incentive change under the parametrization $\theta = u(c, \tilde \omega)/\kappa$ with $(c, \tilde\omega) \notin \{(a, \omega), (b, \omega)\}$ the attention elasticity is zero.
\end{proposition}

Proof. This is an immediate consequence of the invariant likelihood ratio property \citep{caplin2013behavioral} of the Shannon model,
\begin{equation}
    \frac{\gamma^a(\omega)}{\gamma^b(\omega)} = \exp\left(\frac{u(a, \omega) - u(b, \omega)}{\kappa}\right).
\end{equation}
\qed

The restrictions on attention elasticities under the Shannon model is quite strong---the absolute value of elasticity is either one or zero. For instance, in the case where the Shannon model is used in a discrete choice context proposition \ref{prop: elast_shannon} has significant implications for elasticity patterns in the corresponding demand system.

We note however, that typically the scalar $\kappa$ plays a dual role. First, it converts units of information costs (nats) to units of expected utility and allows us to measure all quantities in the same metric. Second, it affects the relative difficulty of acquiring information and hence should affect attention elasticities. With a redefinition of parameters we can consider costs of information where these two roles are treated separately,
\begin{equation}
    K(P, \mu) = \kappa \sum_{\omega, s} P(s\mid\omega)\mu(\omega)\log\left(\frac{P(s\mid \omega) \mu(\omega)}{\mu P_{\mathcal{S}}(s) \mu(\omega)}\right)^\frac{1}{\sigma} = \frac{\kappa}{\sigma} I(P, \mu).
\end{equation}

It is immediate to see that under the above scaled mutual information costs the attention elasticity is constant, $\epsilon^{a, b}_{\omega, \theta} = \sigma$ under the parametrization $\theta = u(a, \omega)/\kappa$. Hence, mutual information implies constant attention elasticity and is the analogue of constant elasticity of substitution (CES) cost functions widely used in production theory. The Shannon model has the distinctive feature that the relative rate of change in posterior odds with changes in incentives is the same irrespective of the initial payoff difference between actions being \$1 or \$1M.

We can gain further insight to this remarkable elasticity property of mutual information costs through the modified logit rule \citep{matvejka2015rational} that the interior conditional choice probabilities satisfy:
\begin{equation}
    P^*(a\mid\omega) \propto \mu P_{\mathcal{S}}^*(a)\exp\left(\frac{u(a,\omega)}{\kappa}\right).
\end{equation}
Conditional choice probabilities are tilted relative to unconditional choice probabilities through an exponential tilting function dependent on the payoff parameters. This link between the barycenter of the optimal experiment, $\mu P^*_{\mathcal{S}}$, and conditional choice probabilities is not unique to the Shannon model. In the following section we introduce a novel class of information cost functions that maintains this relationship while relaxing functional forms.

\section{Information Radius and \texorpdfstring{$\alpha$}{TEXT}-Mutual Information}
\label{sec: info_rad}

Intuitively, without reference to a decision problem, a signal $s$ corresponding to a statistical experiment is informative in as much as the posterior after observing $s$ deviates from the prior. Since the posterior is proportional to the prior times the evidence, $\gamma^s(\omega) \propto \mu(\omega) P(s\mid \omega)$, if the likelihood is constant at $s$ such that $P(s\mid\cdot) = c$ for some $c$, the corresponding posterior coincides with the prior and we obtain no information at $s$.

In fact, most characterizations of information costs can be understood as generalizing how much the likelihoods deviate from a reference constant vector. This requires defining the reference point and a measure of dispersion. In the case of mutual information the reference point is defined by the vector of unconditional signal probabilities, $\mu P_{\mathcal{S}}(s) = \sum_\omega \mu(\omega)P(s\mid\omega)$, while the dispersion is captured by the relative entropy, or KL divergence:
\begin{align}
I(P, \mu) =& \sum_{\omega, s} P(s\mid \omega)\mu(\omega) \log \frac{P(s\mid \omega)\mu(\omega)}{\mu P_\mathcal{S}(s)\mu(\omega)}  \\
=& \sum_\omega \mu(\omega) D\left(P_\omega \Vert \mu P_\mathcal{S}\right). \label{eq: mi2}
\end{align}
As seen by equation \ref{eq: mi2} the mutual information is equal to the average KL divergence between the conditional distributions, $P_\omega$, and the vector of unconditional signal probabilities, $\mu P_\mathcal{S}$.

A useful way to think about the reference point is as the {\it barycenter} of the experiment under the KL divergence and the prior. That is, the reference point is the center of the experiment under the prior in the sense that it is the distribution that minimizes the average KL divergence:
\begin{align}
\mu P_{\mathcal{S}} &= \arg\min_{q\in\Delta(\mathcal{S})} \sum_\omega \mu(\omega) D\left(P_\omega \Vert q\right) \nonumber \\
&= \arg\min_{q\in\Delta(\mathcal{S})} D\left(\mu P \Vert \mu \otimes q\right).
\end{align}
Correspondingly, the mutual information is given by,
\begin{equation}
I(P, \mu) = \min_{q\in\Delta(\mathcal{S})} D\left(\mu P \Vert \mu \otimes q\right).
\end{equation}

We can obtain generalizations of the mutual information by considering different divergence measures and calculating the corresponding deviation from the barycenter. The information radius or $\alpha$-mutual information is a special case of the analogous definition using the R\'{e}nyi divergence \citep{renyi1961measures} instead of the KL divergence.

\begin{definition}[R\'{e}nyi divergence] For two probability distributions, $P, Q$, defined over the same finite sample space $\mathcal{X}$, the R\'{e}nyi divergence of order $\alpha\in (0,1)\cup (1,\infty)$ is defined as,
\begin{equation}
D_\alpha\left(P \Vert Q\right) := \frac{1}{\alpha - 1} \log \left(\sum_{x \in\mathcal{X}}P(x)^\alpha Q(x)^{(1-\alpha)}\right).
\end{equation}
For $\alpha > 1$, we interpret $P(x)^\alpha Q(x)^{(1-\alpha)}$ as $P(x)^\alpha /Q(x)^{(\alpha -1)}$ and adopt the conventions that $0/0 = 0$ and $y/0=\infty$ for $y>0$.
\end{definition}

\begin{definition}[$\alpha$-mutual information] For $\alpha\in (0,1)\cup (1,\infty)$ the $\alpha$-mutual information is defined as,
\begin{align}
I_\alpha\left(P, \mu\right) :=& \min_{q\in\Delta(\mathcal{S})} D_\alpha\left(\mu P \Vert \mu \otimes q\right) \label{eq: a-MI}  \\
=& \min_{q\in\Delta(\mathcal{S})} \frac{1}{\alpha - 1} \log \left(\sum_{\omega, s}\mu(\omega) P(s\mid\omega)^\alpha q(s)^{1-\alpha}\right). \nonumber
\end{align}
\end{definition}

The solution for the minimization problem in \eqref{eq: a-MI}, the barycenter of the experiment, can be expressed in closed form. Suppressing the dependence on $P, \mu$, and $\alpha$, the barycenter is given by,
\begin{equation}
    q^*(s) = \frac{\left(\sum_{\omega}\mu(\omega) P(s\mid \omega)^\alpha \right)^{\frac{1}{\alpha}}}{\sum_{s'} \left(\sum_{\omega}\mu(\omega) P(s'\mid \omega)^\alpha \right)^{\frac{1}{\alpha}}} \qquad \forall s\in\mathcal{S}.
    \label{eq: opt_barycenter}
\end{equation}

Substituting the above expression for the barycenter the $\alpha$-mutual information is given by,
\begin{equation}
I_\alpha\left(P, \mu\right) = \frac{\alpha}{\alpha - 1} \log \sum_{s}\left(\sum_{\omega} \mu(\omega) P(s\mid\omega)^\alpha\right)^{\frac{1}{\alpha}}.
\label{eq: a-CES}
\end{equation}

The information radius was first introduced by \cite{sibson1969information} as a means to generalize measures of dissimilarity among a finite set of weighted probability measures. \cite{verdu2015alpha} pointed out the connection between the information radius and the generalization of mutual information through R\'{e}nyi divergences coining the class \textit{$\alpha$-mutual information}.

\subsection{Properties of \texorpdfstring{$\alpha$}{}-Mutual Information}

It is immediate that for any $\alpha$, $I_\alpha\left(P, \mu\right) \geq 0$ for all $P, \mu$, and is equal to zero for only non-informative experiments---for a non-informative experiment all likelihoods $P(s\mid\cdot)$ are constant and resulting posteriors coincide with the prior. As $\alpha \to 1$ the $\alpha$-mutual information converges to the Shannon mutual information, just as the R\'{e}nyi-divergence converges to the KL divergence. For a fixed prior and experiment the $\alpha$-mutual information is increasing in $\alpha$.\footnote{For further details on $\alpha$-mutual information see the comprehensive reviews of \cite{verdu2015alpha} and \cite{verdu2015conv}.}

Looking at the statistical experiment as a distribution over posteriors provides another potentially illuminating perspective. How much the posterior at a signal $s$ deviates from the prior only depends on the likelihood ratios over states---how different the vector $P(s\mid \cdot)$ is from a constant vector. The cost of an information structure can thus be decomposed to two parts: (i) how much posteriors deviate from the prior, and (ii) what are the probabilities of ending up with different posteriors. With the unconditional signal probabilities being denoted by $\mu P_{\mathcal{S}}$ we can rewrite the cost function as,
\begin{align}
   I_\alpha(P, \mu) &= \frac{\alpha}{\alpha - 1} \log\sum_s\left(\sum_\omega\mu(\omega)P(s\mid\omega)^\alpha\right)^{\frac{1}{\alpha}} \nonumber \\[1em]
   &= \frac{\alpha}{\alpha - 1} \log\sum_s \mu P_{\mathcal{S}}(s)\left(\sum_\omega\mu(\omega)\left(\frac{P(s\mid\omega)}{\mu P_{\mathcal{S}}(s)}\right)^\alpha\right)^{\frac{1}{\alpha}}.
   \label{eq: a_MI_normed}
\end{align}
Since the posterior is as $\gamma^s(\omega) \propto P(s\mid\omega)\mu(\omega)$, constant scaling of the likelihood $P(s\mid\cdot)$ for a given signal $s$ has no impact on the posterior. Hence, from equation \eqref{eq: a_MI_normed} we see that we price the information structure based on the posterior deviating from the prior and the probability with which we obtain the posterior under consideration. The parameter $\alpha$ determines how costly it is to twist the likelihoods away from the constant vector.

Applying the monotone increasing transformation $t\mapsto \frac{1}{\alpha - 1}\exp\left(t\frac{\alpha - 1}{\alpha}\right)$ on the $\alpha$-mutual information we obtain a cost function that is convex in $P$ and that is posterior separable.
    \begin{equation}
        \frac{1}{\alpha - 1} \sum_{s}\left(\sum_{\omega} \mu(\omega) P(s\mid\omega)^\alpha\right)^{\frac{1}{\alpha}} = \frac{1}{\alpha - 1} \sum_{s}\mu P_{\mathcal{S}}(s) \left(\sum_{\omega} \gamma^s(\omega)^\alpha\mu(\omega)^{1 - \alpha}\right)^{\frac{1}{\alpha}}
        \label{eq: post_sep}
    \end{equation}

For any interior $\mu$ the convexity of the map $\phi_\mu\colon \Delta\left(\Omega\right)\to\mathbb{R}$ defined by,
\begin{equation}
    \phi_\mu(\gamma) :=  \frac{1}{\alpha - 1} \left(\sum_{\omega} \gamma(\omega)^\alpha\mu(\omega)^{1 - \alpha}\right)^{\frac{1}{\alpha}},
\end{equation}
implies that the $\alpha$-mutual information is monotone in the Blackwell order.\footnote{For a detailed argument showing Blackwell-monotonicity see appendix \ref{app: blackwell_mon}.} On the other hand, by the results of \cite*{denti2020experimental} we have that the $\alpha$-mutual information is not an experimental cost of information as it approaches zero as the prior becomes more dogmatic. Following \cite{denti2020experimental} it is immediate to form the experimental version of the $\alpha$-mutual information by considering the corresponding information cost under the uniform prior.

As seen from equation \eqref{eq: a-CES} the $\alpha$-mutual information function has the CES property \citep*{arrow1961capital} across states for a given signal---this, however, does not translate to an analogous constant attention elasticity property as remarked in the previous section.

Next, we analyze the class of rational inattention problems under $\alpha$-mutual information costs and refer to the class as $\alpha$-RI.

\section{\texorpdfstring{$\alpha$}{}-RI and Attention Elasticities}
\label{sec: alpha_RI}

We take the information costs to be the scaled $\alpha$-mutual information, $K(P,
\mu) = \kappa I_\alpha(P, \mu)$, and pose the rational inattention problem.
\begin{equation}
    \min_{P\in\mathcal{P}}\  \kappa I_\alpha(P, \mu) - \sum_{\omega, s}\mu(\omega) P(s\mid\omega) u(s, w).
    \label{eq: alpha_RI_direct}
\end{equation}

Since the $\alpha$-mutual information is quasi-convex in $P$ \citep[Theorem 10]{verdu2015conv} and the constraints are convex, the Karush-Kuhn-Tucker (KKT) conditions are necessary and sufficient for characterizing the solution. Denoting the dual variables corresponding to the primal feasibility conditions, $\sum_s P(s\mid\omega) =1, \forall \omega\in\Omega$, by $\lambda_\omega$, and the ones corresponding to the positivity constraints, $P(s\mid\omega) \geq 0, \forall \omega, s \in\Omega\times\mathcal{S}$, by $\delta_{\omega, s}$ the KKT conditions are given by,
\begin{align}
\frac{\alpha}{\alpha-1}&\Bigg(\underbrace{\sum_{\tilde s}\left(\sum_{\tilde \omega} P^*(\tilde s\mid\tilde \omega)^\alpha \mu(\tilde \omega)\right)^{\frac{1}{\alpha}}}_{:= C^*}\Bigg)^{-1}\Bigg(\underbrace{\sum_{\tilde\omega} P^*(s\mid\tilde\omega)^\alpha \mu(\tilde\omega)}_{:= \tilde q^*(s)^{(1-\alpha)}}\Bigg)^{\frac{1-\alpha}{\alpha}} \mu(\omega) P^*(s\mid\omega)^{(\alpha - 1)} \nonumber \\[.9em]
- &\frac{\mu(\omega)u(s, \omega)}{\kappa} + \lambda^*_\omega - \delta^*_{\omega, s} = 0, \label{eq: stationarity_foc} \\[.9em]
&\sum_{s} P^*(s\mid \omega) = 1 \qquad \forall \omega, \\[.9em]
&P^*(s\mid \omega) \geq 0, \quad \delta^*_{\omega, s} \geq 0, \quad P^*(s\mid\omega)\delta^*_{\omega, s} = 0 \qquad \forall \omega, s.
\end{align}

For interior points, $P^*(s\mid\omega) > 0$,
\begin{align}
    &P^*(s\mid\omega) = \left(C^{*^\alpha} \frac{\alpha - 1}{\alpha} \left(\frac{u(s,\omega)}{\kappa} -
    \frac{\lambda^*_\omega}{\mu(\omega)}\right) \right)^{\frac{1}{\alpha -1}}\frac{\tilde q^*(s)}{C^*}.
    \label{eq: interior_FOC}
\end{align}

Note that $q^*(s) := \tilde q^*(s)/ C^*$ is the barycenter of the optimal experiment as given by equation \eqref{eq: opt_barycenter} and is analogous to the unconditional choice probabilities $\mu P_\mathcal{S}^*(s)$ in the Shannon case. Unlike in the Shannon case, we can not express the optimal dual variables $\lambda^*_\omega$ in closed form, however, we can efficiently obtain them numerically as we discuss in section \ref{sec: mod_BA} describing the modified Blahut-Arimoto algortihm.

We state the analogy with the Shannon model explicitly. At interior points, in the Shannon model,
\begin{equation}
    P^*(a\mid\omega) \propto \mu P_{\mathcal{S}}^*(a) \exp\left(\frac{u(a, \omega)}{\kappa} - \frac{\lambda^*_\omega}{\mu(\omega)}\right);
\end{equation}
while in the $\alpha$-RI model,
\begin{equation}
    P^*(a\mid\omega) \propto q^*(a) \left(\frac{\alpha -1}{\alpha}\left(\frac{u(a, \omega)}{\kappa} - \frac{\lambda^*_\omega}{\mu(\omega)}\right)\right)^{\frac{1}{\alpha - 1}}.
\end{equation}

We see that in both cases the optimal conditional choice probabilities are tilted relative to the barycenter of the optimal experiment. In the Shannon model we can express the dual variables in closed form using the primal feasibility conditions,
\begin{equation}
    \lambda^*_\omega = \mu(\omega) \log\left(\sum_{s\in\mathcal{S}} \mu P^*_{\mathcal{S}}(a) \exp\left(\frac{u(s,\omega)}{\kappa}\right)\right).
\end{equation}

For ease of notation we denote,
\begin{equation}
    x^*(s, \omega) := \frac{u(s,\omega)}{\kappa} - \frac{\lambda^*_\omega}{\mu(\omega)},
\end{equation}
which determines the amount by which conditional choice probabilities are tilted relative to the barycenter given the optimal dual variables and payoff parameters.

The restrictive elasticity patterns of the Shannon model are relaxed in $\alpha$-RI problems. The attention elasticities depend on the entire payoff structure and in particular on the baseline payoff difference between the considered actions. To highlight the mechanisms underlying attention elasticities in the $\alpha$-RI model we analyze the class of symmetric tracking problems.

\subsection{Symmetric Problem with Global Incentive Changes}
\label{subsec: sym_global_ch}

We consider \textit{general symmetric tracking problems} defined as follows. Given a finite set of states $n := |\Omega|$ consider the same number of available actions, $\mathcal{A} := \{\ldots, a_{\omega}, \ldots\}$ with $|\mathcal{A}| = n$, such that the payoff structure is given by
\begin{equation}
    u(a, \omega) = \begin{cases}
        h \qquad &\text{if } a = a_\omega, \\
        l \qquad &\text{otherwise},
\end{cases}
\end{equation}
with $h > l$. It is without loss of generality to normalize the low payoff to zero, $l = 0$. Let the prior be uniform over the state space.

Denote $h^* := x^*\left(a_\omega, \omega\right)$ corresponding to action-state pairs that yield high payoff, and $l^* := x^*\left(a_\omega, \tilde\omega\right)$ corresponding to pairs that yield low payoff---note that the dual variables $\lambda^*_\omega$ are constant across $\omega$ due to symmetry.

Arguably, one of the simplest possible change to the incentive scheme is increasing the payoff per unit of information cost for the high-payoff actions in all states simultaneously, $\theta = h/\kappa$. Since the payoff structure is symmetric and the prior is uniform, from the symmetry of the cost function we know that the optimal barycenter, $q^*$, and the unconditional choice probabilities over actions, $\mu P^*_{\mathcal{S}}$, are both uniform over the action/signal space both before and after the change to the incentive scheme. This implies that changes in the posterior ratios are due only to changes in the conditional choice probabilities while unconditional choice probabilities are left intact. That is, in this case the attention elasticity captures the relative rate of change in conditional choice probabilities.

This global incentive change exhibits a ``two-regime'' elasticity pattern as stated by the next proposition.

\begin{proposition}
    \label{prop: elast_global_ch}
    In the general symmetric tracking problem consider any state $\omega$ and two distinct actions, $a_\omega, b$, that are chosen with strictly positive probability conditional on $\omega$ under the optimal strategy of the $\alpha$-RI problem. The attention elasticity under the incentive change $\theta = h/\kappa$ satisfies the following.

    For any $\alpha > 0$,
    \begin{align}
        \epsilon^{a_\omega, b}_{\omega,\theta}\ \begin{cases} \ > 1 \qquad &\text{if } (\alpha - 1)l^* < 1, \\[.9em]
        \ < 1 \qquad &\text{if } (\alpha - 1)l^* > 1.
    \end{cases}
    \end{align}
    Furthermore, for any $\alpha < 1$,
    \begin{equation}
        0 < (\alpha - 1)h^* < \alpha < (\alpha - 1)l^*,
    \end{equation}
    while for $\alpha > 1$,
    \begin{equation}
        0 < (\alpha - 1)l^* < \alpha < (\alpha - 1)h^*.
    \end{equation}
\end{proposition}

Proof. Appendix \ref{app: elast_global_ch}.

Note that the magnitude of $(\alpha - 1)l^*=(1-\alpha)\lambda^*_\omega/\mu(\omega)$ depends on the payoff difference between the high- and low-payoff state-action pairs, $h$. While we don't have a closed-form solution for $h^*$ or $l^*$, by definition we know that their difference is $h^* - l^* = h$. This payoff difference and the curvature of the information cost function determined by $\alpha$ define whether or not the response in attention strategies is elastic or inelastic. In contrast, under the Shannon model the attention elasticity is constant irrespective of the payoff difference.

For cost parameters $\alpha < 1$ the cost of acquiring information is relatively low. When the initial payoff difference is relatively low, the attention elasticity is elastic, that is the relative rate of change in the posterior odds as we increase payoff differences is higher than that under the Shannon case. However, if the initial payoff difference is above a certain threshold the attention elasticity drops below one. We can express the threshold directly in terms of the dual variable, $\lambda^*_\omega$, which is increasing in the payoff difference $h$.

For cost parameters $\alpha > 1$ the cost of acquiring information is relatively high and the elasticity patterns are flipped in an analogous manner. When the initial payoff difference is low,  the attention elasticity is inelastic, that is the relative rate of change in the posterior odds is less than that under the Shannon model. For initial payoff difference above a certain threshold the attention elasticity is greater than 1.

\begin{figure}[htb]
\centering
\begin{subfigure}{.5\textwidth}
  \centering
  \includegraphics[width=.95\linewidth]{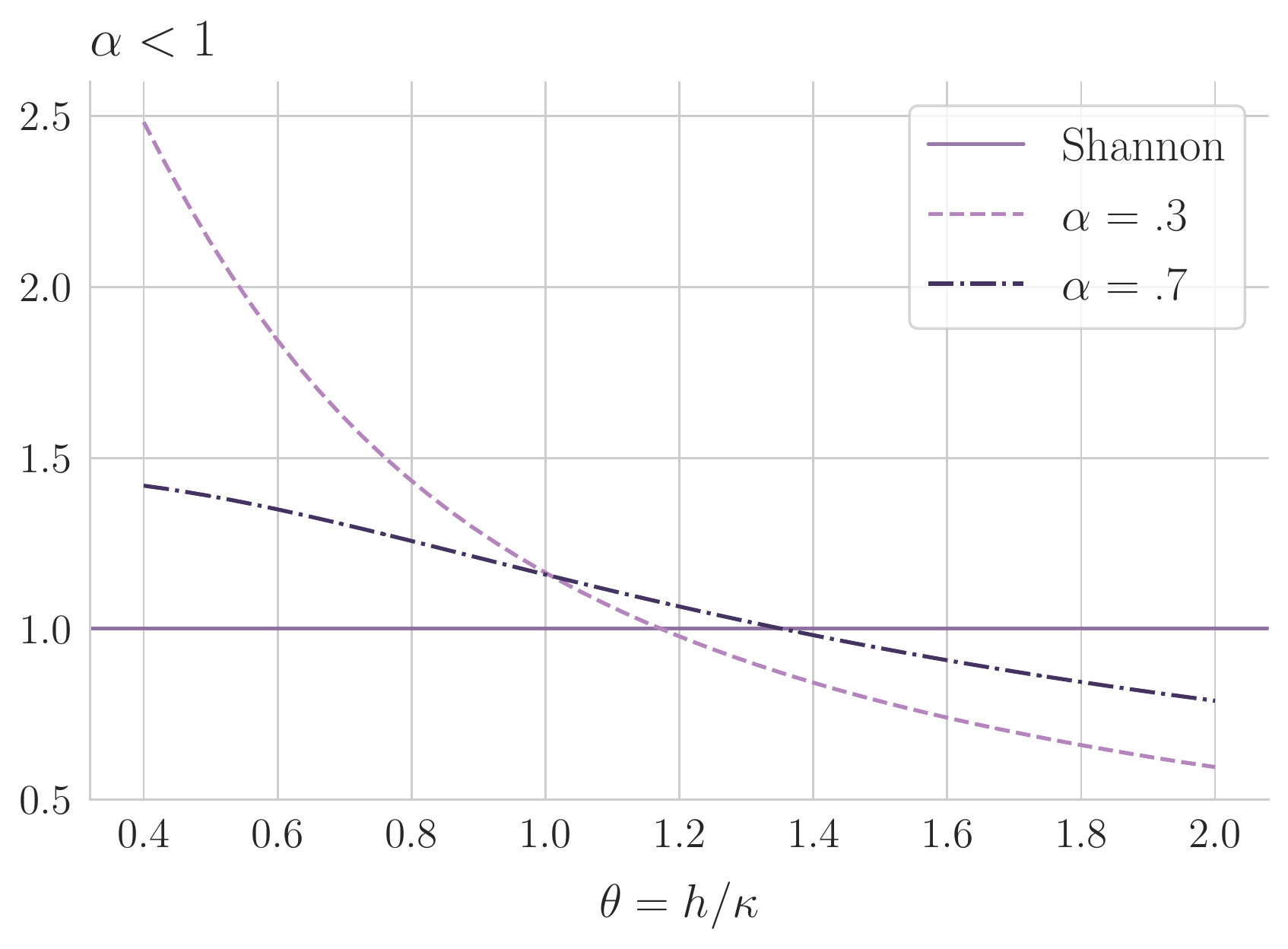}
\end{subfigure}%
\begin{subfigure}{.5\textwidth}
  \centering
  \includegraphics[width=.95\linewidth]{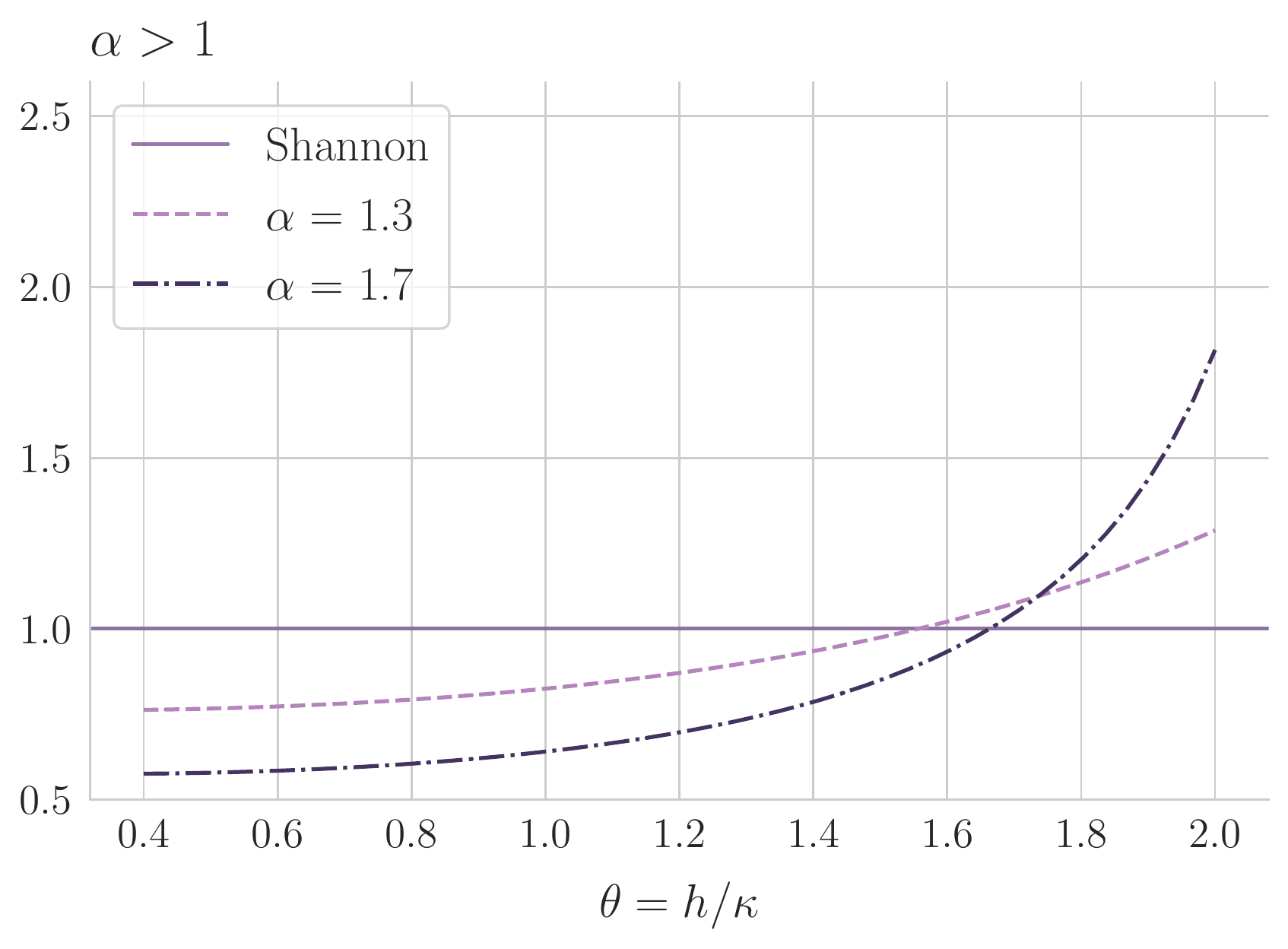}
\end{subfigure}
\caption{Attentional elasticity, $\epsilon^{a_\omega, b}_{\omega,\theta}$}
\label{fig: elast_gl_ch}
\end{figure}

Figure \ref{fig: elast_gl_ch} illustrates the attention elasticity for various levels of initial payoff differences, $h/\kappa$, for different values of the cost parameter. The left-hand-side panel shows attention elasticities when acquiring information is ``easy''---low values of $\alpha$. The right-hand-side panel shows attention elasticities acquiring information is ``hard''.

We require the two actions to be chosen with strictly positive conditional probability to rule out the scenario under $\alpha > 1$ in which the payoff difference is large enough that the optimal attention strategy is the fully informative statistical experiment, under which posterior ratios are not well-defined. Note that this can never happen for $\alpha < 1$ as we discuss in detail in section \ref{sec: ucc}.

\subsection{Invariant Likelihood Ratio Property}

As implied by the first-order conditions for interior points \eqref{eq: interior_FOC} the \textit{invariant likelihood ratio} property \citep{caplin2013behavioral} does not hold under $\alpha$-mutual information costs---except for $\alpha\to1$ as we fall back to the Shannon case.

For signals associated with actions $a$ and $b$ that are chosen with positive probability conditional on a given state under the optimal attention strategy we have,
\begin{equation}
    \frac{P^*(a\mid\omega)}{P^*(b\mid\omega)} = \left(\frac{u(a, \omega)/\kappa -  \lambda^*_\omega/\mu(\omega)}{u(b, \omega)/\kappa - \lambda^*_\omega/\mu(\omega)}\right)^{\frac{1}{\alpha - 1}}\frac{q^*(a)}{q^*(b)}.
\end{equation}

We see that the likelihood ratios are twisted relative to the barycenter ratios by a quantity dependent on the payoff parameters (and the optimal dual variables which also depend on the payoff parameters).

\subsection{Locally Invariant Posteriors}

The \textit{locally invariant posteriors} condition of \cite{caplin2019rationally} states that local changes in the prior---conditional on the set of supported actions not changing---do not lead to changes in the optimal posteriors. \cite{caplin2019rationally} show that locally invariant posteriors is a characteristic feature of uniformly posterior separable cost functions.

The $\alpha$-mutual information is not posterior separable nor uniformly posterior separable and hence the locally invariant posteriors property does not hold. However, as noted in equation \eqref{eq: post_sep} with a monotone increasing transformation we can transform the $\alpha$-mutual information to a posterior separable cost. If we consider the problem of minimizing information costs under the constraint of obtaining a given level of expected utility this transformation should be innocuous (analogous to the expenditure minimization problem).

\section{Invariant Cost Functions}
\label{sec: invariance}

Invariant information costs have important behavioral implications in various economic contexts. In recent work \cite{caplin2019rationally} show that invariance is the defining feature of the Shannon model within the class of uniformly posterior separable cost functions. \cite{hebert2020information} and \cite{angeletos2019inattentive} show that invariant cost functions play a crucial role in the efficiency of equilibria in large games with incomplete information.

While much of the literature has so far focused on generalizations of the Shannon model within posterior separable classes, the above examples demonstrate that invariance can be a desirable feature in many economic contexts. In this section we show that the $\alpha$-mutual information is an invariant class, which is not posterior separable. This provides a novel direction for potential further generalization of the Shannon model.

Invariance imposes two conditions on information cost functions. The first condition---often referred to as information monotonicity---requires that coarsening the information acquired within an event should always weakly reduce information costs. The second condition requires a form of independence from the prior. The least costly way to learn about an event should be invariant to priors that only differ in their relative likelihoods over the event but not in the overall probability mass they place on the event. We follow \cite{caplin2019rationally} and restate the conditions in terms of statistical experiments.

Take a subset of the state space $\bar \Omega \subset \Omega$ and a partition of $\bar \Omega$, $\left\{\bar \Omega_z\right\}_{z\in Z}$. Take any prior $\mu\in \text{int}\Delta(\bar\Omega)$. Next, given experiment $P$ we construct another experiment $P'$ that carries the same information about the events in the partition as $P$ but no information within any event of the partition.

Specifically, given $P$ and $\mu$ define $P'$ as follows:
\begin{enumerate}
    \item[(i)] The marginal signal distributions coincide:
    \begin{equation}
        \text{supp} \mu P'_\mathcal{S} = \text{supp} \mu P_\mathcal{S} \quad \text{and} \quad \mu P'_\mathcal{S}(s) = \mu P_\mathcal{S}(s) \quad \forall s\in \text{supp} \mu P_\mathcal{S}.
        \label{eq: inv1}
    \end{equation}
    \item[(ii)] For each $z\in Z$,
    \begin{itemize}
        \item The corresponding posteriors induced by $P$ and $P'$ put the same mass on events in the partition:
        \begin{equation}
        \sum_{\omega\in\bar\Omega_z} \frac{P(s\mid\omega) \mu(\omega)}{\mu P_\mathcal{S}(s)} = \sum_{\omega\in\bar\Omega_z} \frac{P'(s\mid\omega) \mu(\omega)}{\mu P'_\mathcal{S}(s)}.
        \label{eq: inv2}
        \end{equation}
        \item Conditional on a given event in the partition the posterior induced by $P'$ coincides with the prior. This is equivalent to the likelihood being constant over the event:
        \begin{equation}
            P'(s\mid\omega) = c_z \quad \forall \omega\in\bar\Omega_z.
            \label{eq: inv3}
        \end{equation}
    \end{itemize}
\end{enumerate}

Note that taken together equations \eqref{eq: inv1} and \eqref{eq: inv3}, the condition of \eqref{eq: inv2} implies that the constant likelihood has to satisfy:
\[
c_z =  \frac{\sum_{\omega\in\bar\Omega_z} P(s\mid\omega) \mu(\omega)}{ \sum_{\omega\in\bar\Omega_z} \mu(\omega)}.
\]
It is immediate to check that $\sum_s P'(s\mid\omega) = 1$ for all $\omega\in\bar\Omega$ and all entries are weakly positive hence $P'$ is a valid statistical experiment.

Next take any prior $\bar\mu$ such that the priors put equal weight on events in the partition, i.e. $\bar\mu\left(\bar\Omega_z\right) = \mu\left(\bar\Omega_z\right)$ for all $z\in Z$. Define a statistical experiment $\bar P$ in an analogous way as we defined $P'$ but under $\bar \mu$. Specifically, given $P$, $\mu$, and $\bar \mu$ define $\bar P$ as
\begin{enumerate}
    \item[(i)] The marginal signal distributions coincide:
    \begin{equation}
        \text{supp} \bar\mu \bar{P}_\mathcal{S} = \text{supp} \mu P_\mathcal{S} \quad \text{and} \quad \bar\mu \bar{P}_\mathcal{S}(s) = \mu P_\mathcal{S}(s) \quad \forall s\in \text{supp} \mu P_\mathcal{S}.
        \label{eq: inv1_2}
    \end{equation}
    \item[(ii)] For each $z\in Z$,
    \begin{itemize}
        \item The corresponding posteriors induced by $(P, \mu)$ and $(\bar P, \bar \mu)$ put the same mass on events in the partition:
        \begin{equation}
        \sum_{\omega\in\bar\Omega_z} \frac{P(s\mid\omega) \mu(\omega)}{\mu P_\mathcal{S}(s)} = \sum_{\omega\in\bar\Omega_z} \frac{\bar P(s\mid\omega) \bar \mu(\omega)}{\bar \mu \bar P_\mathcal{S}(s)}
        \label{eq: inv2_2}
        \end{equation}
        \item Conditional on a given event in the partition the posterior induced by $(\bar P, \bar \mu)$ coincides with the prior. This is equivalent to the likelihood being constant over the event:
        \begin{equation}
            \bar P(s\mid\omega) = \bar c_z \quad \forall \omega\in\bar\Omega_z
            \label{eq: inv3_2}
        \end{equation}
    \end{itemize}
\end{enumerate}
Similarly, taken together equations \eqref{eq: inv1_2} and \eqref{eq: inv3_2}, the condition of \eqref{eq: inv2_2} implies that the constant likelihood has to satisfy:
\[
\bar c_z =  \frac{\sum_{\omega\in\bar\Omega_z} P(s\mid\omega) \mu(\omega)}{ \sum_{\omega\in\bar\Omega_z} \bar \mu(\omega)}.
\]

With the above transformations in hand we can define invariant cost functions as in \cite{caplin2019rationally}.

\begin{definition}[Invariance]
A cost function $K$ is invariant if for all finite subsets $\bar\Omega \subset \Omega$, all partitions of $\bar\Omega$, and all pairs $\mu$ and $\bar\mu$ that put equal mass on each event of the partition we have,
\[K\left(P, \mu\right) \geq K\left(P', \mu\right)\]
and
\[K\left(P', \mu\right) = K\left(\bar P, \bar\mu\right).\]
\end{definition}

Our next proposition states that the $\alpha$-mutual information is an invariant cost function.

\begin{proposition}\label{prop: invariance}
With the convention of $\alpha = 1$ denoting the Shannon mutual information, the $\alpha$-mutual information cost function is invariant for any $\alpha > 0$.
\end{proposition}

Proof. Appendix \ref{app: proof_invariance}.

\subsection{Invariance Under Compression}

There are important behavioral implications of the cost function being invariant. If there are events for which all actions that have strictly positive probability of being chosen under the optimal policy also have identical payoffs for each state within the events, then conditional action probabilities are constant within such events under the optimal attention strategy. Formally, suppose there is an event $E$ such that,
\begin{equation}
    u(a, \omega_i) = u(a, \omega_j) \qquad \text{for all } \omega_i, \omega_j \in E, a\in\text{supp} \mu P^*_{\mathcal{S}}.
\end{equation}

Clearly, any two attention strategies, $P$ and $P^\prime$, that satisfy
\begin{align}
\sum_{\omega\in E}\mu(\omega)P(a\mid\omega) &= \sum_{\omega\in E}\mu(\omega)P^\prime(a\mid\omega) \qquad \forall a \in\text{supp} \mu P^*_{\mathcal{S}}, \\[.9em]
P(a \mid \omega) &= P^\prime(a\mid\omega) \qquad \forall \omega\notin E,
\end{align}
yield the same expected utility. By invariance of the $\alpha$-mutual information, among all such attention strategies the one that has minimal cost is such that conditional signal/action probabilities are constant over $E$,
\begin{equation}
    P(a\mid \omega) = c_a \qquad \forall \omega\in E, a \in\text{supp} \mu P^*_{\mathcal{S}}.
\end{equation}
This in turn implies that posteriors over $E$ are proportional to the prior,
\begin{equation}
    \frac{\gamma^a\left(\omega_i\right)}{\mu\left(\omega_i\right)} = \frac{\gamma^a\left(\omega_j\right)}{\mu\left(\omega_j\right)} \qquad \text{for all } \omega_i, \omega_j \in E, a\in\text{supp} \mu P^*_{\mathcal{S}}.
\end{equation}

Invariance under compression states that if two decision problems have the same basic form \citep{caplin2019rationally}, in which we collapse all events like $E$ above, then corresponding attention strategies must coincide. Invariance under compression is the defining feature of the Shannon model \citep{caplin2019rationally} within the class of uniformly posterior separable cost functions. The $\alpha$-mutual information is not posterior separable yet it satisfies invariance under compression.

\begin{proposition}
The solution to the $\alpha$-RI problem satisfies invariance under compression for any cost parameter $\alpha > 0$.
\end{proposition}

Proof. Immediate from the fact that $\alpha$-mutual information class is invariant for all $\alpha > 0$.

\section{The Utility Cost Curve}
\label{sec: ucc}

In \cite*{caplin2020ricspsych} we introduce the \textit{utility cost curve} (UCC) to capture the attention costs of producing different levels of expected utility in a given decision problem. These curves give a succinct visual demonstration of the difficulty of learning in $\alpha$-RI models in relation to the standard Shannon model.

Fixing the payoff structure and the prior, denote the expected utility of decisions made based on experiment $P$ as,
\begin{equation}
    U(P) := \sum_{\omega, a} P(a\mid\omega)\mu(\omega)u(a, \omega).
\end{equation}
The minimum and maximum attainable expected utility levels correspond to the non-informative and perfectly informative experiments, respectively. They are given by,
\begin{align}
    \bar U^{\min} &:= \max_{a\in\mathcal{A}} \sum_\omega \mu(\omega) u(a, \omega); \\[.8em]
    \bar U^{\max} &:= \sum_\omega \mu(\omega) \max_{a\in\mathcal{A}} u(a, \omega).
\end{align}
For any potentially attainable expected utility level $v\in[\bar U^{\min}, \bar U^{\max}]$ the UCC in $\alpha$-RI problems is defined as,
\begin{equation}
    \bar K_\alpha(v) := \min_{P\in\mathcal{P} : U(P) \geq v} I_\alpha(P, \mu).
\end{equation}

Consider the binary symmetric decision problem with uniform prior and payoffs as
\begin{table}[H]
\centering
\begin{tabular}{lcc}
& $a$ & $b$ \\ \hline
$\omega_1$ & 1 & 0 \\
$\omega_2$ & 0 & 1 \\
\end{tabular},
\caption{Binary symmetric problem}
\end{table}
which implies that the ex ante expected utility must be between $1/2$ (no information) and $1$ (perfectly informative experiment).

The utility cost curve of producing expected utility $v\in[1/2, 1]$ under $\alpha$-mutual information costs is
\begin{equation}
    \bar K_\alpha(v) = \frac{\alpha}{\alpha - 1}\log\left(2\left(\frac{1}{2}\left(v^\alpha + (1-v)^\alpha\right)\right)^{\frac{1}{\alpha}}\right),
\end{equation}
which we can derive from the symmetry of the optimal attention strategy.
\begin{figure}[htb]
    \centering
    \includegraphics[width=.65\linewidth]{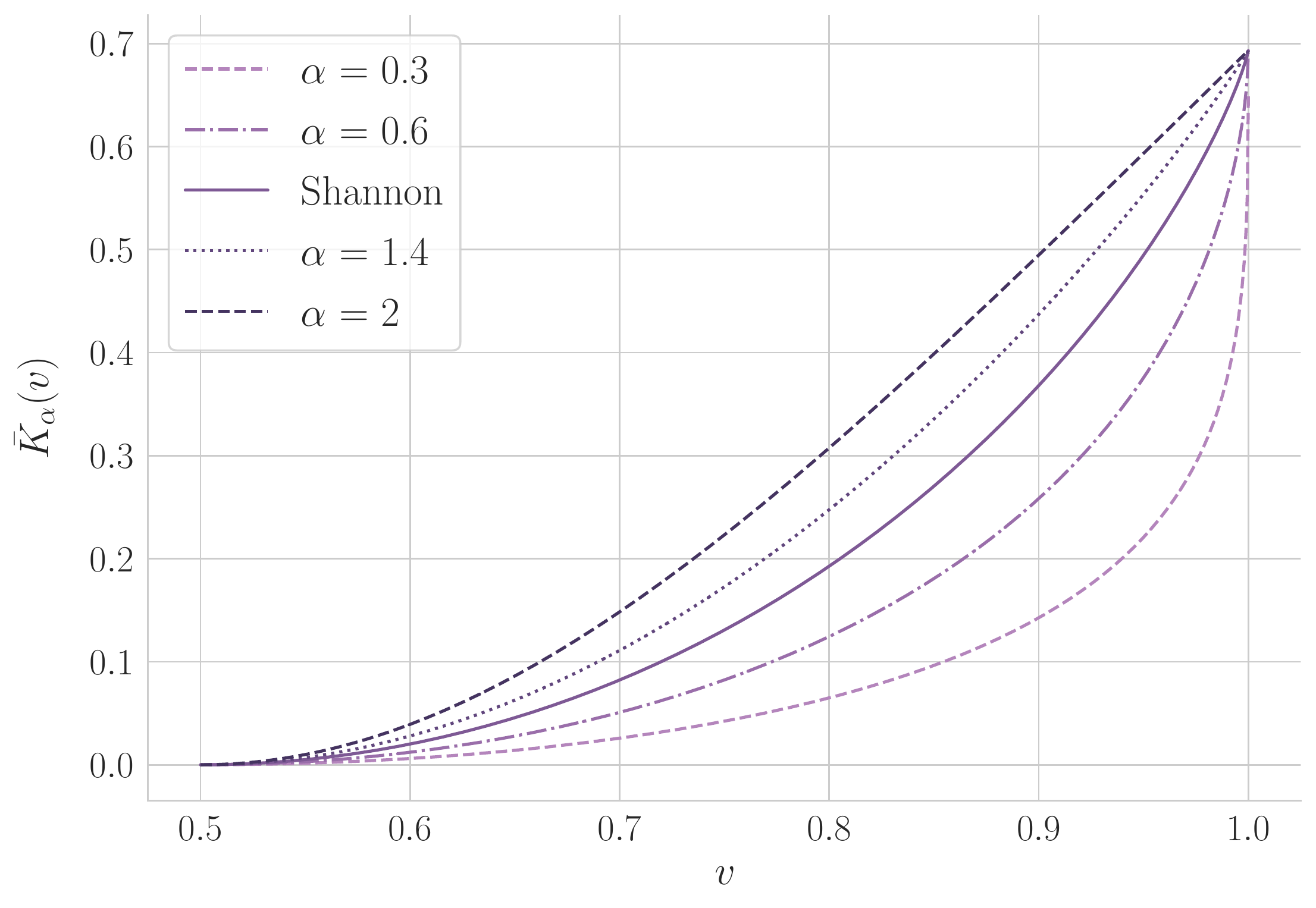}
    \caption{The utility cost curve under different values of $\alpha$}
    \label{fig: ucc}
\end{figure}

Figure \ref{fig: ucc} depicts the utility cost curve for different values of $\alpha$. For each level of attainable expected utility $v\in[1/2, 1]$ the utility cost is increasing in $\alpha$. In general, the cost of the perfectly informative experiment is equal to the R\'{e}nyi entropy of order $1/\alpha$ of the prior, which for a uniform prior of $n$ dimensions is equal to $\log n$ independent of $\alpha$. In the binary symmetric example this implies that the perfectly informative experiments uniformly cost $\log 2$ which we see on figure \ref{fig: ucc} at $v = 1$.

Figure \ref{fig: ucc} shows that as we approach the maximum attentional output, the utility cost curves get steeper for the Shannon case and for $\alpha < 1$. In fact, it is well known that in the Shannon model the DM never chooses an attention strategy that resolves all uncertainty.

Formally, the derivative of the utility cost curve at $v\in\left[1/2, 1\right)$ is,
\begin{equation}
    \bar K_{\alpha}^\prime(v) = \frac{\alpha}{\alpha - 1}\frac{v^{(\alpha-1)} - (1-v)^{(\alpha-1)}}{v^\alpha + (1-v)^\alpha}.
\end{equation}
For any $\alpha < 1$ we have
\[
\lim_{v \to 1} \bar K_{\alpha}^\prime(v) = \infty.
\]
Furthermore, noting that as $\alpha \to 1$  we obtain the Shannon case for which,
\[
\bar K_1(v) = v\log v + (1-v)\log(1-v) + \log 2, \quad \text{and} \quad \bar K_1^\prime(v) = \log\frac{v}{1-v},
\]
we also have $\lim_{v \to 1} \bar K_1^\prime(v) = \infty$. On the other hand, for $\alpha > 1$ we have,
\[
\lim_{v  \to 1}\bar K_{\alpha}^\prime(v) = \frac{\alpha}{\alpha - 1}.
\]
The above features of the cost function imply that in the binary example under $\alpha$-mutual information costs with parameter values $\alpha \leq 1$ a rationally inattentive DM never learns the realization or non-realization of a state with certainty under the optimal attention strategy. This does not necessarily hold for cost parameters $\alpha > 1$. The implications for learning payoff-relevant events with certainty hold in general.

\begin{proposition}
\label{prop: certainty}
For any $\alpha$-RI problem with bounded payoffs we have that:
\begin{itemize}
    \item for $\alpha \leq 1$ there exists no non-trivial event $E \subsetneq \text{supp}(\mu)$ that the DM optimally learns with certainty;
    \item for $\alpha > 1$, and if all payoffs are distinct, there exists a positive scalar $\pi > 0$ such that multiplying all payoffs by $\pi$ the DM optimally learns all events of the form $E = \{\omega : \exists a \text{ s.t. } u(a, \omega) > u(b, \omega)\ \forall b\neq a\}$ with certainty.
\end{itemize}
\end{proposition}

Proof. Appendix \ref{app: proof_certainty}.

Proposition \ref{prop: certainty} states that under Shannon- or $\alpha$-mutual information costs with parameter $\alpha < 1$, a rationally inattentive agent never learns to the point where some of her optimal posteriors put zero weight on states that are in the support of the prior. On the other hand, for $\alpha > 1$ we can always increase the stakes by multiplying the payoffs by a positive scalar such that in the resulting decision problem the DM's optimal posteriors only put weight on events under which a single action is optimal, i.e. for all $a \in \text{supp}(\mu P^*_\mathcal{S})$ we have $\gamma^a(E_a) = 1$ where $E_a := \{\omega : u(a, \omega) > u(b, \omega)\ \forall b\neq a\}$.

\subsection{Incentive-based Psychometric Curve}

From the utility cost curve we can directly obtain the \textit{incentive-based psychometric curve} (IPC) introduced in \cite{caplin2020ricspsych}. This captures the attentional output as we scale the incentives linearly. Formally, fixing the prior and the payoff structure, denote the optimal information structure in the modified $\alpha$-RI problem where all the payoffs are scaled by a positive constant $\pi$ as $P^*_\pi$. We capture the changes in the attentional output as we vary the scaling factor $\pi$ by obtaining the expected utility resulting from $P^*_\pi$ in the original problem. The incentive-based psychometric curve captures the increase in expected utility only due to the changes in the attention strategy but not the increase in payoffs.
\begin{equation}
    \bar U(\pi) := U\left(P^*_\pi \right) = \sum_{a, \omega} P^*_\pi(a\mid\omega)\mu(\omega) u(a, \omega).
\end{equation}

\begin{figure}[htb]
\centering
\begin{subfigure}{.5\textwidth}
  \centering
  \includegraphics[width=.95\linewidth]{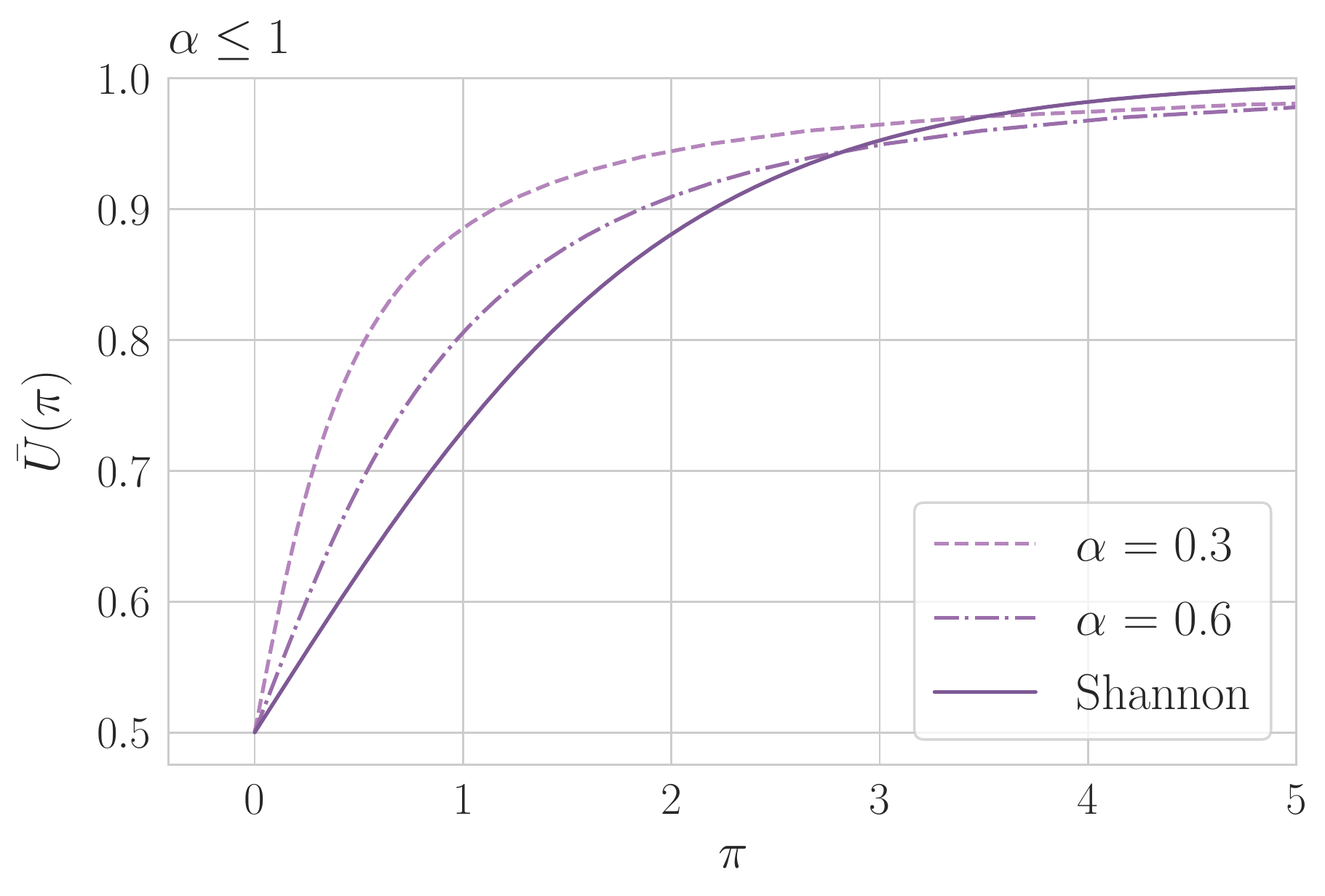}
\end{subfigure}%
\begin{subfigure}{.5\textwidth}
  \centering
  \includegraphics[width=.95\linewidth]{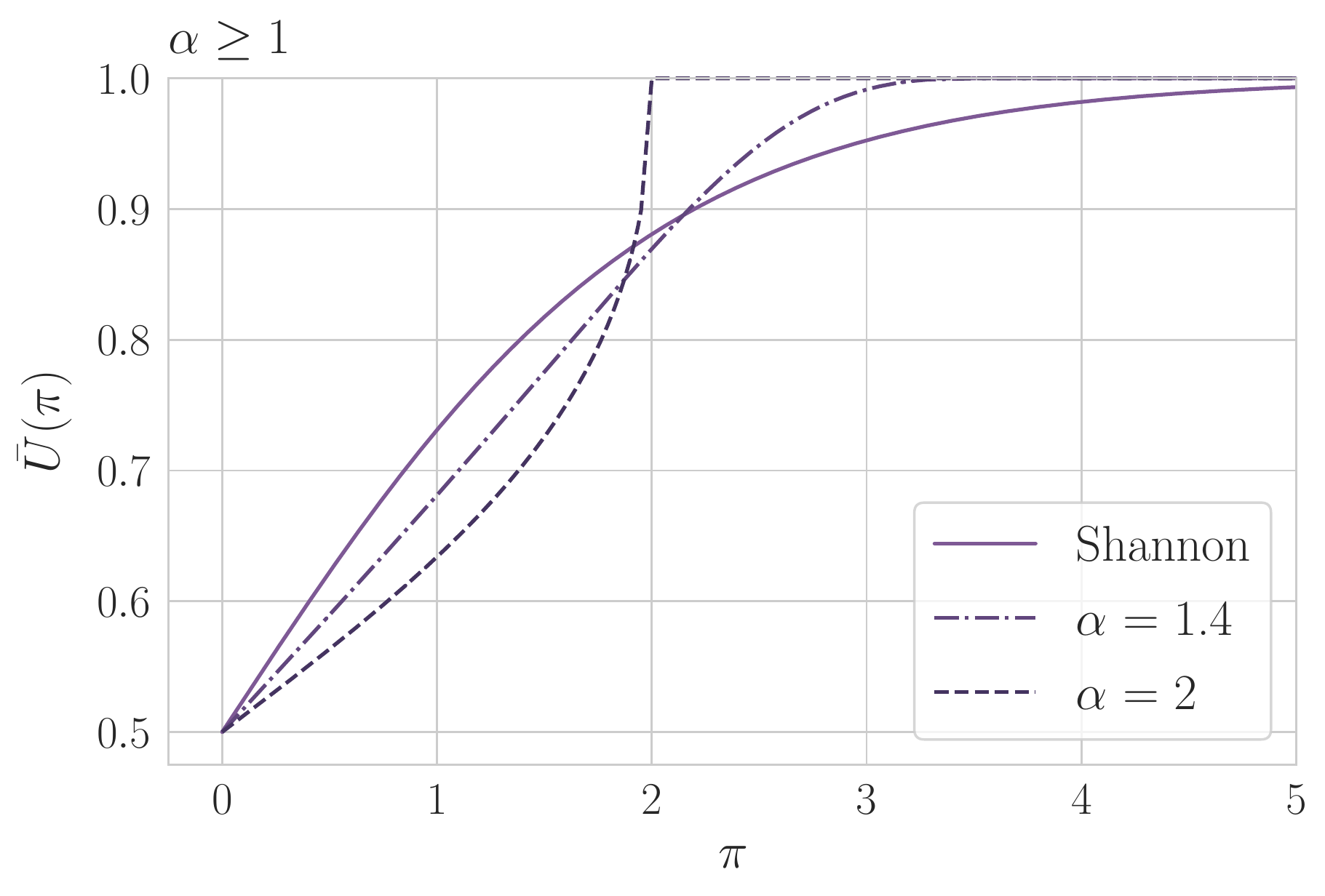}
\end{subfigure}
    \caption{The IPC for different values of $\alpha$}
    \label{fig: ipc}
\end{figure}

As seen in figure \ref{fig: ipc} scaling the payoffs the attentional output of the DM is weakly increasing. For $\alpha \leq 1$ under no circumstances does the DM produce the maximum attentional output, $\bar U^{\max} = 1$. For $\alpha > 1$, there is always a threshold such that scaling the payoffs above the threshold, the DM optimally chooses the perfectly informative experiment and the attentional output attains its maximum in line with the general claims of proposition \ref{prop: certainty}.

\section{Optimal Consideration Sets}
\label{sec: consideration}

Proposition \ref{prop: certainty} characterizes optimal posterior beliefs that put zero probability mass on certain events even though the events are ex ante deemed possible under the prior. The present section characterizes the conditions under which certain actions are never chosen under the optimal attention strategy even though they could be the optimal action under the realization of some states. In this sense, RI models provide a framework to study the formation of optimal consideration sets which arise endogenously dependent on the choice problem. In the Shannon model the formation of optimal consideration sets is analyzed by \cite{caplin2019rational}.

Note that in the $\alpha$-RI model the unconditional choice probability is strictly positive if and only if the corresponding barycenter is strictly positive,
\begin{equation}
    \mu P^*_{\mathcal{S}}(a) > 0 \quad \iff \quad q^*(a) > 0.
\end{equation}
Correspondingly, whether or not an action is in the optimal consideration set can be expressed in terms of the barycenter of the experiment. Denote the function $x \mapsto \max\{x, 0\} =: \left[x \right]^+$ for any $x\in\mathbb{R}$.

\begin{proposition}
    \label{prop: consider}
Under the optimal attention strategy,
\begin{align}
    \frac{\sum_\omega \mu(\omega)\left(\left[\frac{\alpha - 1}{\alpha} \left(\frac{u(a,\omega)}{\kappa} - \frac{\lambda^*_\omega}{\mu(\omega)}\right)\right]^+ \right)^{\frac{\alpha}{\alpha -1}}}{\sum_{\tilde s}q^*(\tilde s) \sum_{\tilde \omega} \mu(\tilde \omega)\left(\left[\frac{\alpha - 1}{\alpha} \left(\frac{u(\tilde s, \tilde \omega)}{\kappa} - \frac{\lambda^*_{\tilde \omega}}{\mu(\tilde \omega)}\right)\right]^+ \right)^{\frac{\alpha}{\alpha -1}}} \leq 1 \qquad \text{for all } a\in\mathcal{A},
\end{align}
with equality for actions that are chosen with strictly positive probability, that is for $a$ such that $\mu P^*_{\mathcal{S}}(a) > 0$. Note that $q^*$ is the barycenter of the optimal experiment $P^*$, while the dual variables $\lambda^*$ are obtained from the primal feasibility constraints on $P^*$.
\end{proposition}

Proof. Appendix \ref{app: proof_consider}.

As we discuss in appendix \ref{app: proof_consider} in detail the conditions characterizing consideration sets can be obtained from the KKT conditions.
The function thresholding at zero, $[\cdot]^+$, appears in the expression to handle corner-solutions under $\alpha > 1$ discussed in the previous section. For cost parameters $\alpha \leq 1$ strictly positive unconditional choice probabilities imply strictly positive conditional choice probabilities by the first claim of proposition \ref{prop: certainty}.
\begin{equation}
    \mu P^*_{\mathcal{S}}(a) > 0 \quad \implies \quad P^*(a\mid \omega) > 0 \quad \forall \omega\in \text{supp} \mu \qquad \left(\text{for } \alpha \leq 1\right).
    \label{eq: interior_a_leq_1}
\end{equation}
However, this is not necessarily the case for cost parameters $\alpha > 1$ as stated by the second claim of proposition \ref{prop: certainty}, which states that the DM might learn some payoff-relevant events with certainty, necessarily contradicting the above condition \eqref{eq: interior_a_leq_1}.

As we show in section \ref{sec: mod_BA} the conditions characterizing the consideration sets are also tightly related to the updating rules in the alternating minimization algorithm that we use to obtain the solution to the $\alpha$-RI problem.

\subsection{Example}

To illustrate the formation of consideration sets consider again the binary decision problem with symmetric payoffs as in section \ref{sec: ucc} under varying the prior probability of state $\omega_1$. As we increase $\mu(\omega_1)$ starting from the uniform prior, action $a$ becomes ex ante optimal with expected utility equal to $\mu(\omega_1)$. The rationally inattentive DM chooses $a$ with higher unconditional probability, while for some threshold value of $\mu(\omega_1)$ the DM exclusively chooses action $a$, that is action $b$ is not in the consideration set.
\begin{figure}[htb]
\centering
\begin{subfigure}{.5\textwidth}
  \centering
  \includegraphics[width=.85\linewidth]{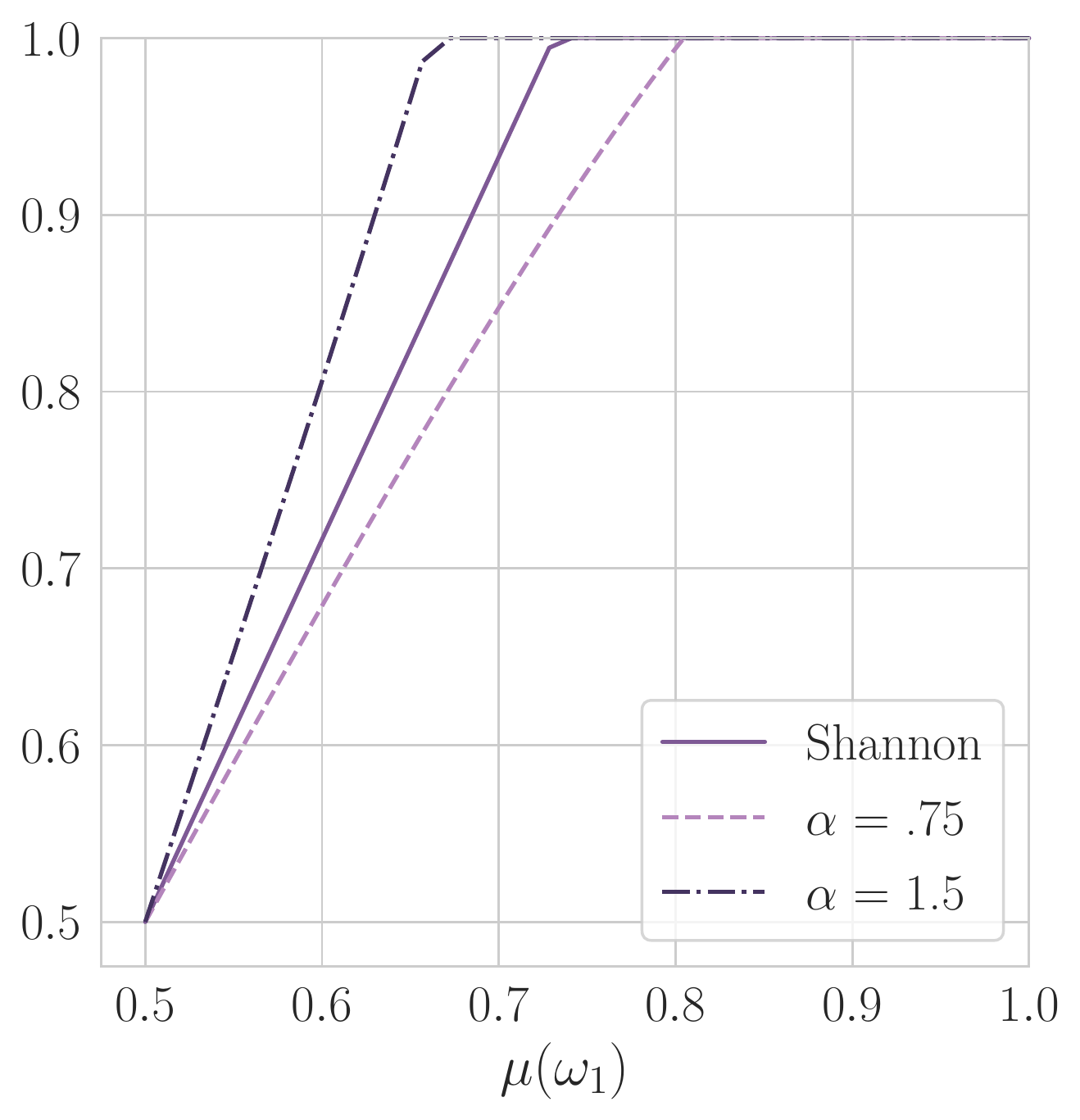}
  \caption{Probability of choosing $a$, $\mu P^*_{\mathcal{S}}(a)$}
  \label{fig: consider_prob}
\end{subfigure}%
\begin{subfigure}{.5\textwidth}
  \centering
  \includegraphics[width=.85\linewidth]{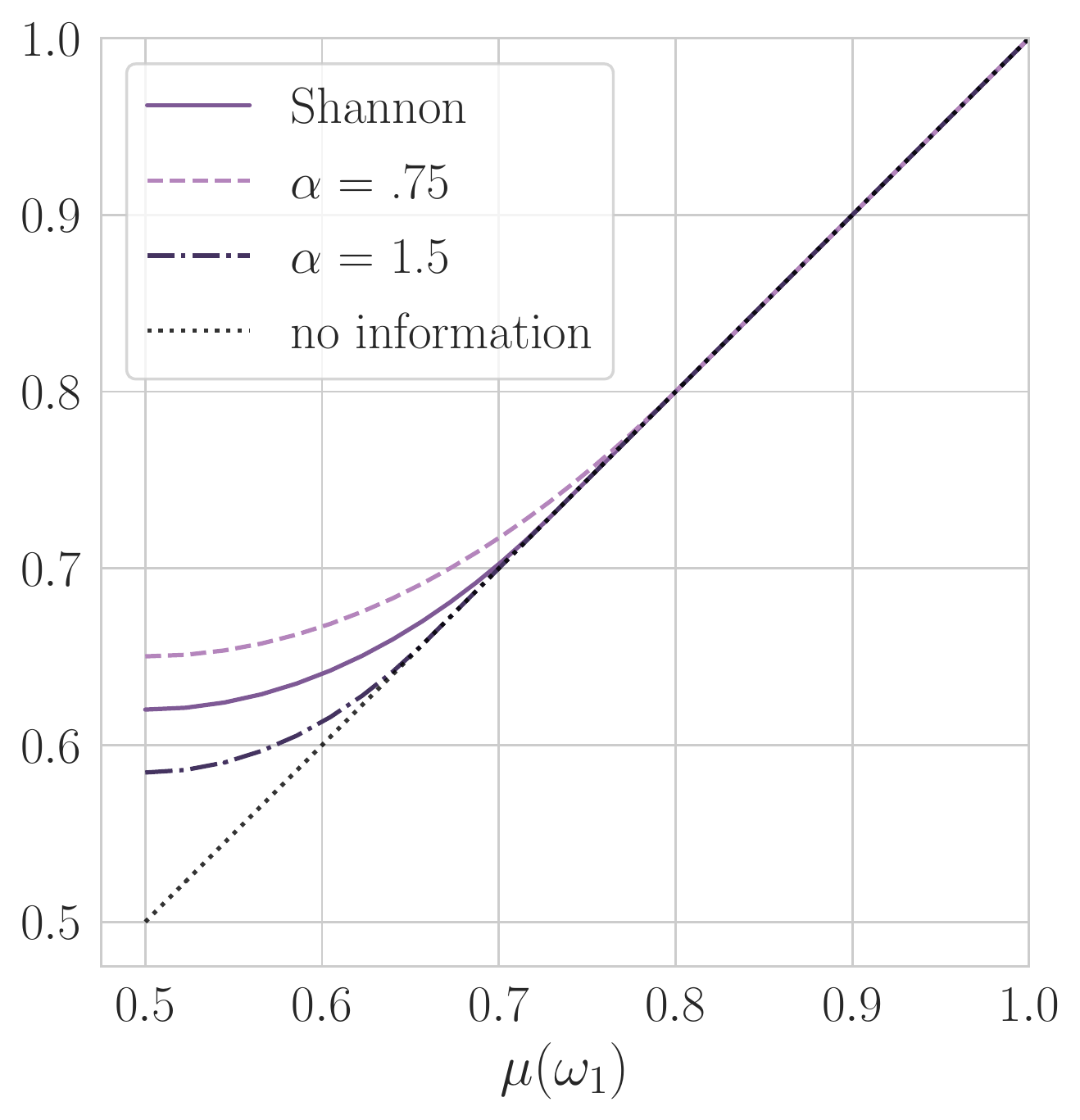}
  \caption{Expected utility net of information costs}
  \label{fig: consider_netU}
\end{subfigure}
\caption{Optimal formation of consideration sets}
\label{fig: consideration_sets}
\end{figure}

Panel \ref{fig: consider_prob} illustrates the unconditional probability of choosing action $a$ as the prior puts more weight on the state in which it is the better action. We see that the higher the value of $\alpha$ the lower the threshold value of the prior $\mu(\omega_1)$ above which the DM exclusively chooses $a$. That is, the costlier the attention the earlier the DM drops actions from the consideration set.

On panel \ref{fig: consider_netU} we see the evolution of the net expected utility the DM achieves relative to the no information benchmark. With only two actions, once the only action that is considered is $a$ there is no information acquisition as the corresponding lines join with the dotted line on the diagonal.

The formation of optimal consideration sets in $\alpha$-RI problems is a natural generalization of the Shannon case where besides beliefs and payoffs the curvature of the information cost function controlled by $\alpha$ determines which actions are chosen under the optimal attention strategy.

\section{The Modified Blahut-Arimoto Algorithm}
\label{sec: mod_BA}

In practice we obtain the solution to the $\alpha$-RI problem through an alternating minimization procedure analogous to the Blahut-Arimoto algorithm \citep{blahut1972,arimoto1972} which is used to obtain solutions for the rational inattention problem under mutual information costs. In the present section we discuss details of implementation and provide proof of convergence.

Anticipating the alternating minimization approach we write the problem using the formulation for the cost function as in equation \eqref{eq: a-MI}.
\begin{align}
    \min_{P\in\mathcal{P}}\  \min_{q\in\Delta(\mathcal{S})}  \kappa D_\alpha(\mu P \Vert \mu\otimes q) - \sum_{\omega, s}\mu(\omega) P(s\mid\omega) u(s, w)
\end{align}

The R\'{e}nyi-divergence is jointly quasi-convex in its two arguments\footnote{For reference see for instance \cite[Theorem 13]{van2014renyi}.} and the constraint sets are both convex so the KKT conditions are necessary and sufficient for obtaining the solution.

Letting $q^0$ be the uniform distribution over all potential signals identified with the action set $\mathcal{S} = \mathcal{A}$ we obtain the sequence $\left\{(P^t, q^t)\right\}_{t\geq 1}$ through alternating minimization,
\begin{align}
    P^t \in\arg\min_{P\in\mathcal{P}} \kappa D_\alpha(\mu P \Vert \mu\otimes q^{t-1}) - \sum_{\omega, s}\mu(\omega) P(s\mid\omega) u(s, w); \\[.9em]
    q^t \in\arg\min_{q\in\Delta(\mathcal{S})} \kappa D_\alpha(\mu P^t \Vert \mu\otimes q) - \sum_{\omega, s}\mu(\omega) P^t(s\mid\omega) u(s, w).
\end{align}

The above alternating-minimization method is known as the block coordinate descent method or Gauss-Seidel method. Its convergence properties are well studied and we can verify the conditions of the general results in \cite{tseng2001convergence} to prove convergence.

\begin{proposition}
    \label{prop: bcd}
    The sequence $\{(P^t, q^t)\}_{t\geq1}$ generated by the alternating minimization algorithm is converging to $(P^*, q^*)$, the critical point of the $\alpha$-RI problem.
\end{proposition}

Proof. Appendix \ref{app: bcd}.

\subsection{Practical Implementation}

Next, we consider details of the modified Blahut-Arimoto algorithm. The algorithm provides and efficient way for obtaining solutions to $\alpha$-RI problems, making their use appealing in applications.\footnote{An implementation of the algorithm in Python is available at \url{https://github.com/danielcsaba/alpha_RI}.}

\subsubsection{Minimization with respect to \texorpdfstring{$q$}{}}

For a given information structure $P^t$ the barycenter of the information structure minimizing the R\'{e}nyi divergence of order $\alpha$ is given by,
\begin{equation}
    q^t(s) = \frac{\left(\sum_{\omega}\mu(\omega) P^t(s\mid \omega)^\alpha \right)^{\frac{1}{\alpha}}}{\sum_{\tilde s} \left(\sum_{\omega}\mu(\omega) P^t(\tilde s\mid \omega)^\alpha \right)^{\frac{1}{\alpha}}}.
    \label{eq: bcd_barycenter}
\end{equation}

For $q^t(s) > 0$ the KKT conditions for the $q$-minimization step imply the above expression. $q^t(s) = 0$ only if $P^t(s\mid\omega) = 0,\  \forall \omega\in\Omega$, hence the expression is valid for cornersolutions as well.

Obtaining the barycenter under the R{\'e}nyi divergence is computationally as efficient as under the KL divergence.

\subsubsection{Minimization with respect to \texorpdfstring{$P$}{}}

Fixing $q^{t-1}$ the KKT conditions characterizing the minimizer $P^t$ are as follows---$\lambda^t$ and $\delta^t$ are the dual variables corresponding to the linear and positivity constraints on the constraint set.
\begin{align}
&\frac{\alpha}{\alpha-1}\Bigg(\underbrace{\sum_{\omega, s} \mu(\omega) P^t(s\mid \omega)^\alpha q^{t-1}(s)^{(1-\alpha)}}_{\text{$:=C^t$}}\Bigg)^{-1}\mu(\omega)\left(\frac{P^t(s\mid\omega)}{q^{t-1}(s)}\right)^{(\alpha - 1)} - \nonumber \\[.9em]
&\frac{\mu(\omega)u(s, \omega)}{\kappa} + \lambda^t_\omega - \delta^t_{\omega, s} = 0 \qquad \forall \omega, s, \\[.9em]
&\sum_{s} P^t(s\mid \omega) = 1 \qquad \forall \omega, \\[.9em]
&P^t(s\mid \omega) \geq 0, \quad \delta^t_{\omega, s} \geq 0, \quad P^t(s\mid\omega)\delta^t_{\omega, s} = 0 \qquad \forall \omega, s
\end{align}

Note that if $q^{t-1}(s) = 0$ then we necessarily have $P^t(s\mid\omega) = 0$ for all $\omega\in\Omega$.

Suppose that $q^{t-1}(s) > 0$.
\begin{itemize}
    \item If $\alpha < 1$ then necessarily $P^t(s\mid\omega) > 0$ for all $\omega\in\Omega$ in line with the first claim of proposition \ref{prop: certainty}. By the complementary slackness condition $\delta^t_{\omega, s} = 0$.
    \item If $\alpha > 1$ and $P^t(s\mid\omega) = 0$ then $\lambda^t_\omega - \frac{\mu(\omega)u(s, \omega)}{\kappa} = \delta^t_{\omega, s} > 0$. Otherwise $\delta^t_{\omega, s} = 0$ by complementary slackness.
\end{itemize}
Taking the above arguments together we have,
\begin{equation}
    P^t(s\mid\omega) = \left(C^t \left[\left(\frac{u(s, \omega)}{\kappa} - \frac{\lambda^t_\omega}{\mu(\omega)}\right)\frac{\alpha - 1}{\alpha}\right]^+ \right)^{\frac{1}{\alpha - 1}} q^{t-1}(s).
\end{equation}
Thresholding the inner term at zero is necessary to take care of the case under $\alpha > 1$ in which $q^{t-1}(s) > 0$ and $P^t(s\mid\omega)=0$. This can only happen if the term $\left(\frac{u(s, \omega)}{\kappa} - \frac{\lambda^t_\omega}{\mu(\omega)}\right)\frac{\alpha - 1}{\alpha} < 0$, so in this case the thresholding function $[\cdot]^+$ returns the correct solution for $P^t(s\mid\omega)$.

While the constant term $C^t$ is a function of $P^t$, it is homogeneous of degree $\alpha$ in $P^t$ if we define $\bar C^t$ by the mapping $P\mapsto \sum_{\omega, s} \mu(\omega) P(s\mid\omega)^\alpha q^{t-1}(s)^{(1-\alpha)}$. We can use this fact to factor it out. Define $P_{-C}^t$ which is proportional to $P^t$ as,
\begin{equation}
     P_{-C^t}^t(s\mid\omega) := \left(\left[\left(\frac{u(s, \omega)}{\kappa} - \frac{\lambda^t_\omega}{\mu(\omega)}\right)\frac{\alpha - 1}{\alpha}\right]^+\right)^{\frac{1}{\alpha - 1}} q^{t-1}(s).
\end{equation}

Then by homogeneity of degree $\alpha$ of $\bar C^t$ we have
\begin{align}
    C^t &= \bar C^t\left(P^t\right) = \bar C^t\left(C^t P^t_{-C^t}\right) = {C^t}^\alpha \bar C^t\left(P^t_{-C^t}\right), \nonumber \\[.8em]
    C^t &= \left(\bar C^t\left(P^t_{-C^t}\right)\right)^{\frac{1}{1 - \alpha}}.
\end{align}

Finally,
\begin{equation}
    P^t(s\mid\omega) = \frac{\left(\left[\left(\frac{u(s, \omega)}{\kappa} - \frac{\lambda^t_\omega}{\mu(\omega)}\right)\frac{\alpha - 1}{\alpha}\right]^+ \right)^{\frac{1}{\alpha - 1}}}{\sum_{\tilde s, \tilde \omega} \mu(\tilde\omega) \left(\left[\left(\frac{u(\tilde s, \tilde \omega)}{\kappa} - \frac{\lambda^t_{\tilde \omega}}{\mu(\tilde \omega)}\right)\frac{\alpha - 1}{\alpha}\right]^+ \right)^{\frac{\alpha}{\alpha - 1}} q^{t-1}(\tilde s)}\ q^{t-1}(s).
    \label{eq: P_update}
\end{equation}

It is instructive to compare this expression with proposition \ref{prop: consider} characterizing consideration sets. Applying the formula for $q^t$ as in \eqref{eq: bcd_barycenter},
\begin{equation}
    q^t(s) = \frac{\left(\sum_\omega \mu(\omega)\left(\left[\left(\frac{u(s, \omega)}{\kappa} - \frac{\lambda^t_\omega}{\mu(\omega)}\right)\frac{\alpha - 1}{\alpha}\right]^+ \right)^{\frac{\alpha}{\alpha - 1}}\right)^{\frac{1}{\alpha}}}{\sum_{\tilde s}\left(\sum_{\tilde \omega} \mu(\tilde\omega) \left(\left[\left(\frac{u(\tilde s, \tilde \omega)}{\kappa} - \frac{\lambda^t_{\tilde \omega}}{\mu(\tilde \omega)}\right)\frac{\alpha - 1}{\alpha}\right]^+ \right)^{\frac{\alpha}{\alpha - 1}}\right)^{\frac{1}{\alpha}} q^{t-1}(\tilde s)}\ q^{t-1}(s).
\end{equation}
We see that either the ratio is one in which case $q^t(s) = q^{t-1}(s)$ or it is less than one in which case $q^t(s)$ converges to zero and the action corresponding to $s$ is not in the consideration set.

\subsubsection{Obtaining Dual Variables}

Equation \eqref{eq: P_update} describes the update step for the information structure given $q^{t-1}$. The expression depends on the dual variables $\lambda^t_\omega$ corresponding to the primal feasibility constraints, $\sum_s P^t(s\mid\omega) = 1, \quad \forall \omega$. Using these conditions we can obtain dual variables $\lambda^t$ through root-finding. To make the dependence on the dual variables explicit write $P^t_\lambda$ and let the primal feasibility conditions be denoted $R^t_\omega(\lambda) := \sum_{s} P^t_\lambda(s\mid\omega) - 1$ where the information structure is defined as in \eqref{eq: P_update} for a candidate $\lambda$. Since we have a closed form expression for the Jacobian we can efficiently apply Newton's method.
\begin{equation}
    \frac{\partial R^t_{\omega_i}}{\partial \lambda_{\omega_k}} = \begin{cases}
        \left(\sum_{s} P^t_\lambda(s\mid\omega_i)\right)\left(\sum_{s} P^t_\lambda(s\mid\omega_k)\right) \qquad &\text{for}\quad i\neq k, \\[1.5em]
        \left(\sum_{s} P^t_\lambda(s\mid\omega_k)\right)^2 - \frac{\sum_{s}\left(\left[\left(\frac{u(s, \omega_k)}{\kappa} - \frac{\lambda_{\omega_k}}{\mu(\omega_k)}\right)\frac{\alpha - 1}{\alpha}\right]^+\right)^{\frac{2 - \alpha}{\alpha - 1}} \frac{q^{t-1}(s)}{\alpha \mu(\omega_k)}} {\sum_{\omega, s} \mu(\omega) \left(\left[\left(\frac{u(s, \omega)}{\kappa} - \frac{\lambda_\omega}{\mu(\omega)}\right)\frac{\alpha - 1}{\alpha}\right]^+\right)^{\frac{\alpha}{\alpha - 1}} q^{t-1}(s)} \qquad &\text{for}\quad i=k.
    \end{cases}
\end{equation}

Furthermore, we have natural bounds on the optimal dual variables depending on the value of $\alpha$ that we enforce in our custom root-finding algorithm. Without enforcing these bounds out-of-the-box root-finding algorithms will typically overshoot and lead to invalid or suboptimal attention strategies. Specifically, in order to satisfy the KKT conditions we need,
\begin{itemize}
    \item for $\alpha < 1$
    \begin{equation}
        \lambda^t_\omega > \max_{a\in\mathcal{A}} \frac{\mu(\omega) u(a, \omega)}{\kappa} \qquad \forall \omega\in\Omega;
        \label{eq: root_bound_l}
    \end{equation}
    \item for $\alpha > 1$
    \begin{equation}
        \lambda^t_\omega <  \max_{a\in\mathcal{A}} \frac{\mu(\omega) u(a, \omega)}{\kappa} \qquad \forall \omega\in\Omega.
        \label{eq: root_bound_h}
    \end{equation}
\end{itemize}

Taken together the modified Blahut-Arimoto algorithm takes the following form:
\begin{figure}[H]
  \centering
  \makebox[.9\linewidth]{
  \begin{minipage}{.9\linewidth}
    \begin{algorithm}[H]
    \SetKwInOut{KwInput}{Input}
    \SetAlgoLined
    \caption{Modified Blahut-Arimoto algorithm}
    \label{algo: mod_BA}
    \KwInput{$(\Omega, \mathcal{A}, u, \mu, \kappa, \alpha)$}
    \KwResult{$P^*, q^*, \lambda^*$}
     $q^0 \leftarrow ones(\vert \mathcal{A}\vert)/\vert \mathcal{A}\vert$\;
     \While{$distance > tol$}{
      $\lambda^t \leftarrow$ via root-finding on primal feasibility given $q^{t-1}$,\\ \hspace{26pt} obeying bounds \eqref{eq: root_bound_l} and \eqref{eq: root_bound_h}\;
      $P^t \leftarrow$ via equation \eqref{eq: P_update} given $q^{t-1}$ and $\lambda^t$\;
      $q^t \leftarrow$ via equation \eqref{eq: bcd_barycenter} given $P^t$\;
      $distance \leftarrow \Vert q^t - q^{t-1} \Vert$
      }
     \end{algorithm}
 \end{minipage}}
\end{figure}

\section{Conclusion}
\label{sec: conclusion}

We consider a generalization of the rational inattention model of \cite{sims1998,sims2003} by measuring costs of information via the information radius of statistical experiments. We show that such costs control elasticities of attention, measuring the sensitivity of attention strategies with changes in incentives. In contrast, the Shannon model restricts attention elasticities to be constant irrespective of the incentives. The resulting class of information costs is invariant but not posterior separable. This provides a novel direction for generalizations of the Shannon model. We explore further behavioral implications relative to the Shannon model and provide an efficient alternating minimization method for obtaining optimal attention strategies.

\pagebreak
\appendix

\bibliographystyle{plainnat}
\bibliography{reference.bib}

\pagebreak
\section{Monotonicity in the Blackwell order}
\label{app: blackwell_mon}

The following Bayesian criterion is a standard characterization of the Blackwell order---for further reference see \cite{blackwell1953equivalent} and \cite{torgersen1991comparison}.

\begin{definition}[Blackwell order]
For two simple statistical experiments, $P, Q$, defined over the same finite state space $\Omega$, experiment $P$ is more informative than experiment $Q$ if for all $\mu\in \text{int}\Delta(\Omega)$ we have,
\begin{equation}
    \sum_s \phi\left(\gamma^s\right)\mu P_{\mathcal{S}}(s) \geq \sum_s \phi\left(\gamma^s\right)\mu Q_{\mathcal{S}}(s),
\end{equation}
for all convex functions $\phi\colon\Delta\left(\Omega\right)\to\mathbb{R}$.
\end{definition}

\begin{proposition}
The $\alpha$-mutual information is monotone in the Blackwell order.
\end{proposition}

Proof. As introduced in equation \eqref{eq: post_sep}, applying the monotone increasing transformation $t\mapsto \frac{1}{\alpha - 1}\exp\left(t\frac{\alpha - 1}{\alpha}\right)$ on the $\alpha$-mutual information we get,
    \begin{equation}
        \frac{1}{\alpha - 1} \sum_{s}\mu P_{\mathcal{S}}(s) \left(\sum_{\omega} \gamma^s(\omega)^\alpha\mu(\omega)^{1 - \alpha}\right)^{\frac{1}{\alpha}}.
    \end{equation}

For monotonicity in the Blackwell order we need to show that the map $\phi_\mu\colon \Delta\left(\Omega\right)\to\mathbb{R}$ defined by,
\begin{equation}
    \phi_\mu(\gamma) :=  \frac{1}{\alpha - 1} \left(\sum_{\omega} \gamma(\omega)^\alpha\mu(\omega)^{1 - \alpha}\right)^{\frac{1}{\alpha}},
\end{equation}
for any  $\mu\in \text{int}\Delta(\Omega)$ is convex.

We show that the Hessian of $\phi_\mu$ is positive semi-definite. Given an enumeration of the finite states denote $\gamma_i = \gamma(\omega_i)$. Given $\alpha, \mu$ and $\gamma$, denote $c := \sum_k \gamma_k^\alpha \mu_k^{1-\alpha}$. The Hessian of $\phi_\mu$ is,
\begin{equation}
    H_{ij} := \frac{\partial^2 \phi_\mu(\gamma)}{\partial \gamma_i\partial \gamma_j} = \begin{cases} -c^{\frac{1}{\alpha}-2} \left(\frac{\gamma_i}{\mu_i}\right)^{\alpha - 1}\left(\frac{\gamma_j}{\mu_j}\right)^{\alpha - 1} \qquad &\text{if} \quad i\neq j; \\[1em]
    -c^{\frac{1}{\alpha}-2} \left(\frac{\gamma_i}{\mu_i}\right)^{2(\alpha - 1)} + c^{\frac{1}{\alpha}-1} \left(\frac{\gamma_i}{\mu_i}\right)^{\alpha - 1}\frac{1}{\gamma_i} \qquad  &\text{if} \quad i = j.
\end{cases}
\end{equation}

For any $x\in\mathbb{R}^{|\Omega|}\setminus \{\mathbf{0}\}$ we have,
\begin{equation}
    x'H x = c^{\frac{1}{\alpha}-1}\left(\sum_j \left(\frac{\gamma_j}{\mu_j}\right)^{\alpha - 1}\frac{1}{\gamma_j}x_j^2  - \frac{1}{c}\left(\sum_j \left(\frac{\gamma_j}{\mu_j}\right)^{\alpha - 1}x_j\right)^2\right).
\end{equation}
As $c$ is strictly positive the sign of $x'Hx$ only depends on the term in the parentheses. Multiplying by $c$ we get,
\begin{equation}
    \left(\sum_j \mu_j^{1-\alpha}\gamma_j^\alpha \left(\frac{x_j}{\gamma_j}\right)^2 \right)\left(\sum_j \mu_j^{1-\alpha}\gamma_j^\alpha \right) - \left(\sum_j \mu_j^{1-\alpha}\gamma_j^\alpha \left(\frac{x_j}{\gamma_j}\right) \right)^2.
\end{equation}

For fixed $\mu, \gamma$, defining the inner product on $\mathbb{R}^{|\Omega|}$ by,
\begin{equation}
    \langle x, y\rangle := \sum_j \mu_j^{1-\alpha}\gamma_j^\alpha x_j y_j,
\end{equation}
we have,
\begin{equation}
    \text{sign}\left(x'Hx\right) = \text{sign}\left(\left\langle \frac{x}{\gamma}, \frac{x}{\gamma}\right\rangle \left\langle \mathbf{1}_{|\Omega|}, \mathbf{1}_{|\Omega|}\right\rangle - \left\langle \frac{x}{\gamma}, \mathbf{1}_{|\Omega|}\right\rangle^2  \right),
\end{equation}
with $\mathbf{1}_{|\Omega|}$ denoting the vector of ones of length $|\Omega|$. By the Cauchy-Schwarz inequality the right hand side is weakly positive, $\phi_\mu$ is positive semi-definite.

\qed

\section{Proof of Proposition \ref{prop: elast_global_ch}}
\label{app: elast_global_ch}

We first derive the formula for the attention elasticity in terms of changes in the dual variables, $\lambda^*$, then analyze the change in the dual via an application of the implicit function theorem. Taking these together we obtain the proof of proposition \ref{prop: elast_global_ch}.

To ease notation---keeping the underlying $\alpha$ parameter fixed---we denote the tilting function,
\begin{align}
    f\colon \mathbb{R} &\to\mathbb{R} \nonumber \\
      x &\mapsto \left(\frac{\alpha - 1}{\alpha} x \right)^{\frac{1}{\alpha -1}}.
\end{align}

$f\left(x^*(s, \omega)\right)$ determines the optimal tilting of the conditional signal probabilities relative to the barycenter,
\begin{equation}
    P^*(s\mid\omega) \propto f\left(x^*(s, \omega)\right) q^*(s).
\end{equation}
At interior points the the quantity $(\alpha - 1)x^*(s, \omega)) > 0$ so the function $f$ is well-defined. We deal with boundary points in later sections.

The posterior ratio for a fixed state under the $\alpha$-mutual information is given by,
\begin{equation}
    \Gamma^{a, b}_\omega = \frac{\gamma^a(\omega)}{\gamma^b(\omega)} = \frac{f\left(x^*(a, \omega)\right)}{f\left(x^*(b, \omega)\right)}\frac{\sum_{\tilde \omega} \mu(\tilde\omega)f\left(x^*(b, \tilde\omega)\right)}{\sum_{\tilde \omega} \mu(\tilde\omega)f\left(x^*(a, \tilde\omega)\right)}.
\end{equation}

With these notations in hand the general formula for the attention elasticity is given by,
\begin{align}
    \epsilon^{a_\omega,b}_{\omega, \theta} = &\frac{f^\prime\left(x^*(a_\omega, \omega)\right){x^*}^\prime(a_\omega, \omega)}{f\left(x^*(a_\omega, \omega)\right)} - \frac{f^\prime\left(x^*(b, \omega)\right){x^*}^\prime(b, \omega)}{f\left(x^*(b, \omega)\right)} + \nonumber \\[.8em]
    &\frac{\sum_{\tilde\omega}\mu(\tilde\omega) f^\prime\left(x^*(b, \tilde \omega)\right){x^*}^\prime(b, \tilde \omega)}{\sum_{\tilde\omega}\mu(\tilde\omega) f\left(x^*(b, \tilde\omega)\right)} - \frac{\sum_{\tilde\omega}\mu(\tilde\omega) f^\prime\left(x^*(a_\omega, \tilde \omega)\right){x^*}^\prime(a_\omega, \tilde \omega)}{\sum_{\tilde\omega}\mu(\tilde\omega) f\left(x^*(a_\omega, \tilde\omega)\right)},
    \label{eq: elast_gen_form}
\end{align}
with,
\begin{align}
    f^\prime(x) &:= \frac{\mathrm{d} f}{\mathrm{d} z}\Big|_{z=x} = \frac{1}{\alpha}\left(\frac{\alpha -1}{\alpha}x\right)^{\frac{2-\alpha}{\alpha - 1}}, \\[.8em]
    {x^*}^\prime(s, \omega) &:= \frac{\partial x^*}{\partial \theta} = \begin{cases}
        1 - \frac{{\lambda^*}^\prime_\omega}{\mu(\omega)} \qquad &\text{if } s = s_\omega, \\
        - \frac{{\lambda^*}^\prime_\omega}{\mu(\omega)}\qquad &\text{otherwise}.
\end{cases}
\end{align}
The derivative of the dual is with respect to the incentive change $\theta$,
\[
{\lambda^*}^\prime_\omega := \frac{\partial \lambda^*_\omega}{\partial \theta},
\]
and we write $D_\theta \boldsymbol{\lambda^*}$ to denote the vector of derivatives for all states.

Note that in the general symmetric tracking problem with the global incentive change the last two term of equation \ref{eq: elast_gen_form} representing changes in the unconditional choice probabilities cancel out---as the unconditional choice probabilities stay uniform.

Note further that,
\begin{equation}
    \frac{f^\prime(x)}{f(x)} = \frac{1}{(\alpha - 1)x}.
\end{equation}

Correspondingly we have,
\begin{equation}
    \epsilon^{a_\omega,b}_{\omega, \theta} = \frac{{\lambda^*}^\prime_\omega}{\mu(\omega)}\left(\frac{1}{(\alpha - 1)x^*(b, \omega)} - \frac{1}{(\alpha - 1)x^*(a_\omega, \omega)}\right) + \frac{1}{(\alpha - 1)x^*(a_\omega, \omega)}.
\end{equation}

We derive the change in the dual variable via an application of the implicit function theorem on the primal feasibility constraints. We analyze this under a general problem with potentially asymmetric incentive structure.

\subsection{Expressing Changes in the Dual Variables}

The primal feasibility constraints are given by\footnote{We can use the first-order conditions \eqref{eq: interior_FOC} and the fact that $C^*$ is homogeneous of degree 1 in the information structure.
},
\begin{equation}
    R^*_\omega := \frac{\sum_{s}f\left(x^*(s,\omega)\right)q^*(s)}{\left(\sum_{s}\left(\sum_{\tilde\omega}f\left(x^*(s,\tilde\omega)\right)^\alpha\mu(\tilde\omega)\right)^{\frac{1}{\alpha}}q^*(s)\right)^\alpha} = 1 \qquad \forall\omega\in\Omega.
\end{equation}

To ease notation we denote the constant in the denominator,
\[
\xi := \left(\sum_{s}\left(\sum_{\tilde\omega}f\left(x^*(s,\tilde\omega)\right)^\alpha\mu(\tilde\omega)\right)^{\frac{1}{\alpha}}q^*(s)\right)^\alpha,
\]
and note that for any $\omega\in\Omega$ the primal feasibility implies,
\[
\sum_{s}f\left(x^*(s,\omega)\right)q^*(s) = \xi.
\]

To make the dependence on the parameter of interest, $\theta$, explicit we rewrite the vector of primal feasibility constraints as,
\[
\boldsymbol{R^*}(\theta, \boldsymbol{\lambda^*}(\theta), \boldsymbol{q^*}(\theta)) = \boldsymbol{1}.
\]

By the implicit function theorem, we have that,
\begin{equation}
    D_\theta \boldsymbol{\lambda^*} = \left[D_{\boldsymbol{\lambda^*}}\boldsymbol{R^*}\right]^{-1}_{|\Omega|\times|\Omega|}\left(-\left[D_{\boldsymbol{q^*}}\boldsymbol{R^*}\right]_{|\Omega|\times|\mathcal{S}|} D_\theta \boldsymbol{q^*} -  D_{\theta}\boldsymbol{R^*} \right).
\end{equation}

We can exploit the structure of the Jacobian with respect to $\boldsymbol{\lambda^*}$ to obtain the inverse. Note that,
\begin{equation}
    D_{\boldsymbol{\lambda^*}}\boldsymbol{R^*}_{\omega, \tilde\omega} = \begin{cases}
    1 \qquad &\text{if } \omega \neq \tilde\omega, \\
    1 - \frac{1}{\mu(\omega)}\frac{\sum_s f^\prime(x^*(s,\omega)) q^*(s)}{\xi}\qquad &\text{if } \omega = \tilde\omega.
\end{cases}
\end{equation}
Since all off-diagonal elements are constant we can apply the Sherman-Morrison formula to obtain the inverse of the Jacobian,
\begin{equation}
    \left[D_{\boldsymbol{\lambda^*}}\boldsymbol{R^*}\right]^{-1} =
    I_{|\Omega|}\boldsymbol{d} - \frac{1}{1 + \langle \boldsymbol{d}, \boldsymbol{1}_{|\Omega|}\rangle}\boldsymbol{d}\boldsymbol{d}',
\end{equation}
where $I_{|\Omega|}$ is the identity matrix of size $|\Omega|\times|\Omega|$, $\boldsymbol{1}_{|\Omega|}$ is the vector of ones of length $|\Omega|$, and $\boldsymbol{d}$ is the vector with entries,
\begin{equation}
    \boldsymbol{d}_\omega = -\frac{\mu(\omega)\xi}{\sum_s f^\prime(x^*(s,\omega)) q^*(s)}.
\end{equation}

Further,
\begin{equation}
    D_{\boldsymbol{q^*}}\boldsymbol{R^*}_{\omega, s} = \frac{f\left(x^*(s, \omega)\right)}{\xi} - A,
\end{equation}
for some constant $A$. Since $D_\theta \boldsymbol{q^*}$ sums to zero we can ignore $A$.

Analogous derivations pertaining to the implicit function theorem hold for arbitrary incentive structures.

Continuing with the proof of the current section for the case of the general symmetric tracking problem with $\theta = h/\kappa$, denote $n := |\Omega|$ and
\begin{equation}
    x^*(s, \omega) = \begin{cases}
        h^*\qquad &\text{if } s = a_\omega, \\
        l^*\qquad &\text{otherwise}.
\end{cases}
\end{equation}

We have,
\begin{equation}
    D_{\theta}\boldsymbol{R^*} = \boldsymbol{1}_{|\Omega|}\frac{\left(f^\prime\left(h^*\right) - f\left(h^*\right)\right)q^*(a_\omega)}{\xi}.
\end{equation}

Note that in the current case $D_\theta \boldsymbol{q^*} = \boldsymbol{0}$ and $\boldsymbol{d}_\omega$ is constant across $\omega$ so we have,
\begin{equation}
    \left[D_{\boldsymbol{\lambda^*}}\boldsymbol{R^*}\right]^{-1}\boldsymbol{1}_{|\Omega|} = \boldsymbol{1}_{|\Omega|} \frac{\xi}{\left(f\left(h^*\right) - f^\prime\left(h^*\right)\right) + \left(n - 1\right)\left(f\left(l^*\right) - f^\prime\left(l^*\right)\right)}.
\end{equation}

Combining we have,
\begin{equation}
    {\lambda^*}^\prime_\omega = \frac{\left(f\left(h^*\right) - f^\prime\left(h^*\right)\right)q^*(a_\omega)}{\left(f\left(h^*\right) - f^\prime\left(h^*\right)\right) + \left(n - 1\right)\left(f\left(l^*\right) - f^\prime\left(l^*\right)\right)}.
\end{equation}

We can express the difference,
\begin{equation}
    f(x) - f^\prime(x) = f(x)\left(1- \frac{1}{(\alpha - 1)x}\right)
\end{equation}

Noting that both $q^*$ and $\mu$ are uniform we arrive at,
\begin{align}
    \epsilon^{a_\omega,b}_{\omega, \theta} &= \frac{f(h^*)\left(1- \frac{1}{(\alpha - 1)h^*}\right)\left(\frac{1}{(\alpha - 1)l^*} - \frac{1}{(\alpha - 1)h^*}\right)}{f(h^*)\left(1- \frac{1}{(\alpha - 1)h^*}\right) + (n-1)f(l^*)\left(1- \frac{1}{(\alpha - 1)l^*}\right)} + \frac{1}{(\alpha - 1)h^*} \nonumber \\[1em]
    &= \frac{f(h^*)\left(1- \frac{1}{(\alpha - 1)h^*}\right)\frac{1}{(\alpha-1)l^*}  + (n-1)f(l^*)\left(1 - \frac{1}{(\alpha - 1)l^*}\right)\frac{1}{(\alpha-1)h^*}}{f(h^*)\left(1- \frac{1}{(\alpha - 1)h^*}\right) + (n-1)f(l^*)\left(1- \frac{1}{(\alpha - 1)l^*}\right)}
    \label{eq: eps_magn}
\end{align}

We next prove a lemma on the magnitudes of $(\alpha - 1)h^*$ and $(\alpha - 1)l^*$ and then finish the proof.

\begin{lemma}
    \label{lemma: magnitude}
    In the general symmetric tracking problem for $\alpha < 1$ the optimal solution satisfies,
    \begin{equation}
        (\alpha - 1)h^* < \alpha < (\alpha - 1)l^*;
    \end{equation}
    for $\alpha > 1$ the optimal interior` solution satisfies,
    \begin{equation}
        (\alpha - 1)h^* > \alpha > (\alpha - 1)l^*.
    \end{equation}
\end{lemma}

Proof. First, take the stationarity first-order condition \eqref{eq: stationarity_foc} and rearrange,
\begin{equation}
    \frac{\left( \sum_{\tilde \omega} P^*\left(s\mid \tilde \omega\right)^\alpha \mu(\tilde \omega)\right)^{-1}}{\left(\sum_{\tilde s}\left(\sum_{\tilde \omega} P^*\left(\tilde s\mid \tilde \omega\right)^\alpha \mu(\tilde \omega)\right)^{\frac{1}{\alpha}}\right)^{\frac{\alpha}{\alpha -1}}}  P^*(\omega)^\alpha = \left(\frac{\alpha -1}{\alpha}\left(\frac{u(s, \omega)}{\kappa} - \frac{\lambda^*_\omega}{\mu(\omega)} + \frac{\delta^*_{\omega, s}}{\mu(\omega)}\right)\right)^{\frac{\alpha}{\alpha - 1}}.
\end{equation}
Multiplying by $\mu(\omega)$ and summing across all states yields,
\begin{align}
    \Psi := &\left(\sum_{\tilde s}\left(\sum_{\tilde \omega} P^*\left(\tilde s\mid \tilde \omega\right)^\alpha \mu(\tilde \omega)\right)^{\frac{1}{\alpha}}\right)^{\frac{\alpha}{1 - \alpha}} = \nonumber \\[1em]
    &\sum_\omega \mu(\omega) \left(\frac{\alpha -1}{\alpha}\left(\frac{u(s, \omega)}{\kappa} - \frac{\lambda^*_\omega}{\mu(\omega)} + \frac{\delta^*_{\omega, s}}{\mu(\omega)}\right)\right)^{\frac{\alpha}{\alpha - 1}}\quad \forall s\in\mathcal{S}.
    \label{eq: foc_constant}
\end{align}
Note that the above condition holds for all actions $s$ irrespective of whether or not they are chosen with strictly positive probability under the optimal strategy. The constant $\Psi$ is strictly positive and real---it is equal to $\Psi = \exp\left(-I_\alpha(P^*, \mu)\right)$.

Using the primal feasibility constraints and noting that as $q^*$ is uniform $\xi$ is equal to the right-hand side of equation $\eqref{eq: foc_constant}$ we have that,
\begin{align}
    \sum_{s}f\left(x^*(s,\omega)\right)q^*(s) &= \xi = \exp\left(-I_\alpha(P^*, \mu)\right) \leq 1 \qquad \forall \omega\in\Omega; \label{eq: unif_bary} \\[.9em]
    \sum_{\omega}f\left(x^*(s,\omega)\right)^\alpha\mu(\omega) &= \xi = \exp\left(-I_\alpha(P^*, \mu)\right) \leq 1 \qquad \forall s \ \text{with}\ q^*(s) > 0.
\end{align}
We obtain the inequalities by the fact that the $\alpha$-mutual information is always weakly positive, $I_\alpha(P^*, \mu) \geq 0$.

In the symmetric problem the uniform barycenter $q^*$ in \eqref{eq: unif_bary} implies,
\begin{align}
    \frac{1}{n}f\left(h^*\right) + \frac{n-1}{n}f\left(l^*\right) \leq 1 \nonumber \\
    f\left(l^*\right)\left(\underbrace{\frac{f\left(h^*\right)}{f\left(l^*\right)}}_{>1} + (n-1)\right) \leq n
    \label{eq: exp_I}
\end{align}

Next, use the equality,
\begin{equation}
    \sum_{\tilde s}f\left(x^*(\tilde s,\omega)\right)q^*(\tilde s) = \sum_{\tilde \omega}f\left(x^*(s,\tilde \omega)\right)^\alpha\mu(\tilde \omega) \quad \text{for any } \omega\ \text{and }  s \ \text{with}\ q^*(s) > 0,
\end{equation}
which in the symmetric tracking problem implies,
\begin{equation}
    \underbrace{\left(\frac{\alpha - 1}{\alpha}h^*\right)^{\frac{\alpha}{\alpha - 1}}}_{> 0}
    \left(1 - \left(\frac{\alpha - 1}{\alpha}h^*\right)^{-1}\right) = \underbrace{(n-1)\left(\frac{\alpha - 1}{\alpha}l^*\right)^{\frac{\alpha}{\alpha - 1}}}_{> 0}
    \left(\left(\frac{\alpha - 1}{\alpha}l^*\right)^{-1} - 1\right).
    \label{eq: signs}
\end{equation}

If $\alpha < 1$ then necessarily $(\alpha - 1)l^* > (\alpha - 1)h^* > 0$ since $l < h$ and conditional probabilities are interior. This implies $f\left(h^*\right) > f\left(l^*\right) > 0$ and the term in the bracket of equation \eqref{eq: exp_I} is greater than $n$. To satisfy the equation we need
\begin{equation}
    f\left(l^*\right) < 1 \quad \implies \quad \frac{\alpha - 1}{\alpha}l^* > 1.
\end{equation}
And to match the signs of the two sides of equation \eqref{eq: signs} we need,
\begin{equation}
    \frac{\alpha - 1}{\alpha}h^* < 1.
\end{equation}

If $\alpha > 1$ then necessarily $(\alpha - 1)h^* > (\alpha - 1)l^* > 0$ since $h < l$ and conditional probabilities are interior. This implies $f\left(h^*\right) > f\left(l^*\right) > 0$ and the term in the bracket of equation \eqref{eq: exp_I} is greater than $n$. To satisfy the equation we need
\begin{equation}
    f\left(l^*\right) < 1 \quad \implies \quad \frac{\alpha - 1}{\alpha}l^* \leq 1.
\end{equation}
And to match the signs of the two sides of equation \eqref{eq: signs} we need,
\begin{equation}
    \frac{\alpha - 1}{\alpha}h^* > 1.
\end{equation}
\qed

Lastly, note that equation \eqref{eq: signs} implies that under $\alpha < 1$ the denominator of equation \eqref{eq: eps_magn} is always negative, while under $\alpha > 1$ it is always positive. 

To complete the proof of proposition \ref{prop: elast_global_ch} we use lemma \ref{lemma: magnitude} and equation \eqref{eq: eps_magn}.

\textbf{Case I.} $\alpha < 1$.

By lemma \ref{lemma: magnitude} we have $(\alpha - 1)l^* > \alpha > (\alpha - 1)h^* > 0$ and $f\left(h^*\right) > f\left(l^*\right) > 0$. Given the denominator of \eqref{eq: eps_magn} being negative,
\begin{align}
    \epsilon^{a_\omega,b}_{\omega, \theta} > 1 \qquad \text{if}\quad (\alpha - 1)l^* < 1, \nonumber \\[.8em]
    \epsilon^{a_\omega,b}_{\omega, \theta} < 1 \qquad \text{if}\quad (\alpha - 1)l^* > 1.
\end{align}

\textbf{Case II.} $\alpha > 1$.

By lemma \ref{lemma: magnitude} we have $(\alpha - 1)h^* > \alpha > (\alpha - 1)l^* > 0$ and $f\left(h^*\right) > f\left(l^*\right) > 0$. Given the denominator of \eqref{eq: eps_magn} being positive,
\begin{align}
    \epsilon^{a_\omega,b}_{\omega, \theta} > 1 \qquad \text{if}\quad (\alpha - 1)l^* < 1, \nonumber \\[.8em]
    \epsilon^{a_\omega,b}_{\omega, \theta} < 1 \qquad \text{if}\quad (\alpha - 1)l^* > 1.
\end{align}

\qed

\section{Proof of Proposition \ref{prop: invariance}}
\label{app: proof_invariance}

Note that the $\alpha$-mutual information for $\alpha\in (0,1)\cup (1,\infty)$ is given by,
\begin{equation}
I_\alpha\left(P, \mu\right) = \frac{\alpha}{\alpha - 1} \log \sum_{s}\left(\sum_{\omega} \mu(\omega) P(s\mid\omega)^\alpha\right)^{\frac{1}{\alpha}}.
\end{equation}

First, we prove information monotonicity. With the transformations introduced in section \ref{sec: invariance} defining $P^\prime, \bar P$, and $\bar \mu$, we need to show,
\[I_\alpha\left(P, \mu\right) \geq I_\alpha\left(P^\prime, \mu\right).\]

Take the monotone increasing transformation $\phi \colon t\mapsto \frac{1}{\alpha - 1}\exp\left(t\frac{\alpha - 1}{\alpha}\right)$, information monotonicity is equivalent to,
\begin{align*}
    \phi\left(I_\alpha\left(P, \mu\right)\right) &\geq \phi\left(I_\alpha\left(P^\prime, \mu\right)\right) \\[.9em]
    \frac{1}{\alpha - 1} \sum_{s}\left(\sum_{\omega} \mu(\omega) P(s\mid\omega)^\alpha\right)^{\frac{1}{\alpha}} &\geq \frac{1}{\alpha - 1} \sum_{s}\left(\sum_{\omega} \mu(\omega) P^\prime(s\mid\omega)^\alpha\right)^{\frac{1}{\alpha}}.
\end{align*}
Note that the transformed $\alpha$-mutual information is posterior separable as previously remarked in \eqref{eq: post_sep}.

Given that likelihoods of $P^\prime$ are constant over elements of the partition we can write,
\begin{align}
    \phi\left(I_\alpha\left(P^\prime, \mu\right)\right) &= \frac{1}{\alpha - 1} \sum_{s}\left(\sum_{z\in Z} \left(c_z^\alpha \sum_{\omega\in\bar\Omega_z} \mu(\omega)\right) \right)^{\frac{1}{\alpha}}
    \label{eq: inv_proof3} \\[.9em]
    &= \frac{1}{\alpha - 1} \sum_{s}\left(\sum_{z\in Z} \left(\sum_{\omega\in\bar\Omega_z}P(s\mid\omega) \mu(\omega)\right)^\alpha \left(\sum_{\omega\in\bar\Omega_z} \mu(\omega)\right)^{(1-\alpha)} \right)^{\frac{1}{\alpha}}.
    \label{eq: inv_proof1}
\end{align}

On the other hand,
\begin{equation}
\phi\left(I_\alpha\left(P, \mu\right)\right) = \frac{1}{\alpha - 1} \sum_{s}\left(\sum_{z\in Z} \sum_{\omega\in\bar\Omega_z}\left(P(s\mid\omega) \mu(\omega)\right)^\alpha \mu(\omega)^{(1-\alpha)} \right)^{\frac{1}{\alpha}}.
\label{eq: inv_proof2}
\end{equation}

Denote the inner function in the parentheses as $f_\alpha(x, y) := x^\alpha y^{(1-\alpha)}$. Note that $f_\alpha(x, y)$ is the ``conic transform'' of $\tilde f_\alpha(x) := x^\alpha$, that is $f_\alpha(x, y) = \tilde f_\alpha(x/y)y$ for $y>0$. Since the prior is $\mu\in\text{int}\Delta\left(\bar\Omega\right)$ we have no issues applying the transformation. Next, we will use the fact that the conic transform preserves convexity/concavity.

We have two cases to consider. If $\alpha < 1$,
\begin{itemize}
    \item $\tilde f_\alpha$ is concave, so the conic transform $f_\alpha$ is also concave. By Jensen's inequality we have,
    \[\frac{1}{|\bar \Omega_z|} \sum_{\omega\in\bar\Omega_z}\left(P(s\mid\omega) \mu(\omega)\right)^\alpha \mu(\omega)^{(1-\alpha)} \leq \frac{1}{|\bar \Omega_z|} \left(\sum_{\omega\in\bar\Omega_z}P(s\mid\omega) \mu(\omega)\right)^\alpha \left(\sum_{\omega\in\bar\Omega_z} \mu(\omega)\right)^{(1-\alpha)}. \]
    \item The inequality holds for all $z\in Z$ so it is preserved for the sum over the partitions.
    \item Since the mapping $x\mapsto x^{\frac{1}{\alpha}}$ is monotone increasing for $x > 0$ and the scalar $\frac{1}{\alpha - 1} < 0$, we have that information monotonicity holds,
    \[I_\alpha\left(P, \mu\right) \geq I_\alpha\left(P', \mu\right).\]
\end{itemize}

For $\alpha > 1$ we can apply an analogous argument with $\tilde f_\alpha$ being convex, and so $f_\alpha$ also being convex. Applying Jensen's inequality and noting that $\frac{1}{\alpha - 1} > 0$ we have,
\[I_\alpha\left(P, \mu\right) \geq I_\alpha\left(P', \mu\right).\]

Next we show that
\[I_\alpha\left(P', \mu\right) = I_\alpha\left(\bar P, \bar\mu\right).\]

Note, that since by construction $\mu\left(\bar\Omega_z\right) = \bar\mu\left(\bar\Omega_z\right)$ for all $z\in Z$ we have that,
\[
c_z = \frac{\sum_{\omega\in\bar\Omega_z} P(s\mid\omega) \mu(\omega)}{ \sum_{\omega\in\bar\Omega_z}  \mu(\omega)} = \frac{\sum_{\omega\in\bar\Omega_z} P(s\mid\omega) \mu(\omega)}{ \sum_{\omega\in\bar\Omega_z} \bar \mu(\omega)} = \bar c_z \qquad \forall z\in Z.
\]
Now use the decomposition as in equation \eqref{eq: inv_proof3},
\begin{align*}
\phi\left(I_\alpha\left(P', \mu\right)\right) &= \frac{1}{\alpha - 1} \sum_{s}\left(\sum_{z\in Z} \left(c_z^\alpha \sum_{\omega\in\bar\Omega_z} \mu(\omega)\right) \right)^{\frac{1}{\alpha}} \\[.9em]
&= \frac{1}{\alpha - 1} \sum_{s}\left(\sum_{z\in Z} \left(\bar c_z^\alpha \sum_{\omega\in\bar\Omega_z}\bar\mu(\omega)\right) \right)^{\frac{1}{\alpha}} = \phi\left(I_\alpha\left(\bar P, \bar \mu\right)\right).
\end{align*}
This completes our proof. \qed

\section{Proof of Proposition \ref{prop: certainty}}
\label{app: proof_certainty}

The proof for the claims of proposition \ref{prop: certainty} hinges on the directional derivatives of the $\alpha$-mutual information evaluated at points on the boundary of the constraint set.

Note that the partial derivative of $I_\alpha$ with respect to $P(a\mid \omega)$ is
\begin{equation}
    \frac{\partial I_\alpha(P, \mu)}{\partial P(a\mid\omega)} = \frac{\alpha}{\alpha - 1}\frac{\left(\sum_{\tilde \omega}\mu(\tilde\omega)P(a\mid\tilde \omega)^\alpha\right)^{\frac{1-\alpha}{\alpha}}}{\sum_{\tilde s}\left(\sum_{\tilde \omega}\mu(\tilde \omega)P(\tilde s\mid\tilde \omega)^\alpha\right)^{\frac{1}{\alpha}}}\mu(\omega)P(a\mid\omega)^{\alpha -1}.
    \label{eq: marginal_a_MI}
\end{equation}

We treat the two claims separately.

Claim 1. For $\alpha \leq 1$ there  exists no non-trivial event $E \subsetneq \text{supp}(\mu)$ that the DM optimally learns with certainty.

Suppose there exists such a set. Then necessarily there exists $a\in\mathcal{A}$ such that $P^*(a\mid \omega_i) = 0$ for some $i$ while $P^*(a\mid \omega_j) > 0$ for some $j$. Furthermore, since signal probabilities have to add up to one, there exists some $b$ such that $P^*(b\mid \omega_i) > 0$. Moving in the feasible direction that results in a new experiment $(P^*(a\mid \omega_i) + \epsilon, P^*(b\mid \omega_i) - \epsilon)$ with all other conditional signal probabilities being the same as in $P^*$ reduces information costs by infinity at the margin as shown by equation \eqref{eq: marginal_a_MI} while only affecting payoffs by a finite amount. This contradicts the optimality of $P^*$.

Claim 2. For $\alpha > 1$ take an event of the form $E_a := \left\{\omega : u(a,\omega) > u(b, \omega) \ \forall b\neq a\right\}$. Suppose that the optimal attention strategy under the scaled payoff, $\pi u(\cdot, \cdot)$, is such that $\gamma^a\left(E_a\right) < 1$. This implies that $P^*(a\mid \omega_k) > 0$ for some $\omega_k\notin E_a$. Since by assumption all payoffs are distinct, $\omega_k \in E_b$ for some other action $b\neq a$ which implies $u(b, \omega_k) > u(a, \omega_k)$. Moving in the feasible direction that results in a new experiment $(P^*(b\mid\omega_k) + \epsilon, P^*(a\mid\omega_k) - \epsilon)$ with all other conditional signal probabilities being the same as in $P^*$ changes information costs by a finite amount on the margin as shown by equation \eqref{eq: marginal_a_MI}, so there must be $\pi$ such that the considered move has greater utility increase on the margin $\pi (u(b, \omega_k) - u(a, \omega_k))$. This would contradict the optimality of the information structure inducing $\gamma^a\left(E_a\right) < 1$.

\qed

\section{Proof of Proposition \ref{prop: consider}}
\label{app: proof_consider}

First, take the stationarity first-order condition \eqref{eq: stationarity_foc} and rearrange,
\begin{equation}
    \frac{\left( \sum_{\tilde \omega} P^*\left(s\mid \tilde \omega\right)^\alpha \mu(\tilde \omega)\right)^{-1}}{\left(\sum_{\tilde s}\left(\sum_{\tilde \omega} P^*\left(\tilde s\mid \tilde \omega\right)^\alpha \mu(\tilde \omega)\right)^{\frac{1}{\alpha}}\right)^{\frac{\alpha}{\alpha -1}}}  P^*(\omega)^\alpha = \left(\frac{\alpha -1}{\alpha}\left(\frac{u(s, \omega)}{\kappa} - \frac{\lambda^*_\omega}{\mu(\omega)} + \frac{\delta^*_{\omega, s}}{\mu(\omega)}\right)\right)^{\frac{\alpha}{\alpha - 1}}.
\end{equation}
Multiplying by $\mu(\omega)$ and summing across all states yields,
\begin{align}
    \Psi := &\left(\sum_{\tilde s}\left(\sum_{\tilde \omega} P^*\left(\tilde s\mid \tilde \omega\right)^\alpha \mu(\tilde \omega)\right)^{\frac{1}{\alpha}}\right)^{\frac{\alpha}{1 - \alpha}} = \nonumber \\[1em]
    &\sum_\omega \mu(\omega) \left(\frac{\alpha -1}{\alpha}\left(\frac{u(s, \omega)}{\kappa} - \frac{\lambda^*_\omega}{\mu(\omega)} + \frac{\delta^*_{\omega, s}}{\mu(\omega)}\right)\right)^{\frac{\alpha}{\alpha - 1}}\quad \forall s\in\mathcal{S}.
\end{align}
Note that the above condition holds for all actions $s$ irrespective of whether or not they are chosen with strictly positive probability under the optimal strategy. The constant $\Psi$ is strictly positive and real.

We consider two cases depending on the value of $\alpha$.

\textbf{Case 1. } $\alpha < 1$.

If $\mu P^*_{\mathcal{S}}(a) > 0$ (or equivalently $q^*(a) > 0$) then $P^*(a\mid \omega) > 0$ for all $\omega$ by proposition $\ref{prop: certainty}$. By the complementary slackness conditions we have $\delta^*_{\omega, a} = 0$ for all $\omega$. Hence we have,
\begin{equation}
    \sum_\omega \mu(\omega) \left(\frac{\alpha -1}{\alpha}\left(\frac{u(s, \omega)}{\kappa} - \frac{\lambda^*_\omega}{\mu(\omega)}\right)\right)^{\frac{\alpha}{\alpha - 1}} = \Psi \quad \forall s \quad \text{such that } q^*(s) > 0.
\end{equation}

If $\mu P^*_{\mathcal{S}}(a) = 0$ (or equivalently $q^*(a) = 0$) then $P^*(a\mid \omega) = 0$ for all $\omega$. By the complementary slackness and dual feasibility conditions $\delta^*_{\omega, a} > 0$. Since $\alpha -1 < 0$ we have that,
\begin{equation}
    \sum_\omega \mu(\omega) \left(\frac{\alpha -1}{\alpha}\left(\frac{u(s, \omega)}{\kappa} - \frac{\lambda^*_\omega}{\mu(\omega)}\right)\right)^{\frac{\alpha}{\alpha - 1}} < \Psi \quad \forall s \quad \text{such that } q^*(s) = 0.
\end{equation}
Note that irrespective of $P^*(s\mid\omega)$ being strictly positive $\frac{\alpha -1}{\alpha}\left(\frac{u(s, \omega)}{\kappa} - \frac{\lambda^*_\omega}{\mu(\omega)}\right)$ is positive otherwise the stationarity condition would be violated.

\textbf{Case 2. } $\alpha > 1$.

Under $\alpha > 1$ we might have $P^*(a\mid\omega) = 0$ while $\mu P^*(a) > 0$. Consider such a case. By the stationarity first-order condition we have,
\begin{equation}
    \frac{u(a, \omega)}{\kappa} - \frac{\lambda^*_\omega}{\mu(\omega)} = -\frac{\delta^*_{\omega, a}}{\mu(\omega)},
\end{equation}
and since $\alpha - 1>0$,
\begin{equation}
    \left[\frac{\alpha -1}{\alpha}\left(\frac{u(s, \omega)}{\kappa} - \frac{\lambda^*_\omega}{\mu(\omega)}\right)\right]^+ = \frac{\alpha -1}{\alpha}\left(\frac{u(s, \omega)}{\kappa} - \frac{\lambda^*_\omega}{\mu(\omega)} + \frac{\delta^*_{\omega, s}}{\mu(\omega)}\right).
\end{equation}
Correspondingly we arrive at,
\begin{equation}
    \sum_\omega \mu(\omega) \left(\left[\frac{\alpha - 1}{\alpha}\left(\frac{u(s, \omega)}{\kappa} - \frac{\lambda^*_\omega}{\mu(\omega)}\right)\right]^+\right)^{\frac{\alpha}{\alpha - 1}} = \Psi \quad \forall s \quad \text{such that } q^*(s) > 0.
\end{equation}
Note that without thresholding the expression at zero for the terms where $P^*(s\mid \omega) = 0$ would render the exponentiation ill-defined as we would be taking the fractional power of a negative number.

If $\mu P^*_{\mathcal{S}}(a) = 0$ (or equivalently $q^*(a) = 0$) then $P^*(a\mid \omega) = 0$ for all $\omega$. By the complementary slackness and dual feasibility conditions $\delta^*_{\omega, a} > 0$. Since $\alpha - 1 > 0$ we have that,
\begin{equation}
    \sum_\omega \mu(\omega) \left(\left[\frac{\alpha - 1}{\alpha}\left(\frac{u(s, \omega)}{\kappa} - \frac{\lambda^*_\omega}{\mu(\omega)}\right)\right]^+\right)^{\frac{\alpha}{\alpha - 1}} < \Psi \quad \forall s \quad \text{such that } q^*(s) = 0.
\end{equation}

Now taking the convex combination of these expressions with the convex weights given by the barycenter of the optimal experiment,
\begin{equation}
    \sum_{\tilde s}q^*(\tilde s) \sum_{\tilde \omega} \mu(\tilde \omega)\left(\left[\frac{\alpha - 1}{\alpha} \left(\frac{u(\tilde s, \tilde \omega)}{\kappa} - \frac{\lambda^*_{\tilde \omega}}{\mu(\tilde \omega)}\right)\right]^+ \right)^{\frac{\alpha}{\alpha -1}} = \Psi.
\end{equation}
By the fact that $\mu P^*_{\mathcal{S}}(s) > 0 \iff q^*(s) > 0$ proposition \ref{prop: consider} holds if we substitute the unconditional signal probabilities for the barycenter.
\qed

\section{Proof of Proposition \ref{prop: bcd}}
\label{app: bcd}

Since the objective is bounded from below and the objective function is decreasing at each step the algorithm is converging. That does not necessarily imply that the limit point is the global optimum of the problem. We can use the fact that algorithm \ref{algo: mod_BA} is a special case of the block non-linear Gauss-Seidel method---also called block coordinate descent method---with quasi-convex objective function, two blocks, and convex constraints. We can directly apply theorem 5.1 of \cite{tseng2001convergence}.

The conditions that we need to verify in order to apply the results of \cite{tseng2001convergence} are:
\begin{itemize}
    \item The R\'{e}nyi divergence, $D_\alpha$ is continuous on its effective domain, which holds for finite distributions.
    \item $D_\alpha$ is quasi-convex in both arguments---in fact it is jointly quasi-convex \citep[Theorem 13]{van2014renyi}---and hemivariate in both arguments.
    \item $D_\alpha$ is lower semi-continuous \citep[Theorem. 15]{van2014renyi}.
    \item The effective domain of $D_\alpha$ is a product of subsets of Euclidean spaces.
\end{itemize}

The only property that we need to show is that $D_\alpha$ is hemivariate in both arguments separately on its effective domain. A function is hemivariate in an argument if it is not constant on any line segment belonging to the effective domain. That is for a given $Q$ take distinct $P_0$ and $P_1$ such that the line segment $\lambda (P_0, Q) + (1 - \lambda)(P_1, Q)$ for any $\lambda\in[0,1]$ is in the effective domain of $D_\alpha$. Suppose that $D_\alpha$ is not hemivariate. Then with $P_\lambda := \lambda P_0 + (1 - \lambda)P_1$,
\begin{equation}
    D_\alpha(P_\lambda\Vert Q) := \frac{1}{\alpha - 1}\log \left(\sum_{x\in\mathcal{X}} P_\lambda(x)^\alpha Q(x)^{(1-\alpha)}\right) = C \qquad  \forall \lambda\in[0,1].
\end{equation}

We can derive a contradiction by considering
\begin{equation}
    \sum_{x\in\mathcal{X}} P_\lambda(x)^\alpha Q(x)^{(1-\alpha)} = \exp \left((\alpha - 1)C\right) \qquad  \forall \lambda\in[0,1],
\end{equation}
and using the strict convexity or strict concavity of the function $z\mapsto z^\alpha$ depending on the value of $\alpha$. The case for $Q$ is analogous.

As a result theorem 5.1 of \cite{tseng2001convergence} applies and the modified Blahut-Arimoto algorithm converges to a stationary point of the objective function at which all KKT conditions are satisfied.

\qed

\end{document}